\documentclass[12pt,preprint]{aastex}
\newcommand{\kms}{{km~s$^{-1}$}}
\newcommand{\cm}{cm$^{-2}$}

\shorttitle{Evolution of atomic gas and 21~cm \ion{H}{1} absorption}
\shortauthors{Braun}

\begin{document}


\title{Cosmological evolution of atomic gas and implications for 21~cm \ion{H}{1} absorption} 


%%\author{R. Braun\altaffilmark{1}}
\author{Robert Braun}
\affil{CSIRO Astronomy and Space Science, PO Box 76, Epping, NSW 1710, Australia}
\email{Robert.Braun@csiro.au}



\begin{abstract}
Galaxy disks are shown to contain a significant population of atomic
clouds of 100~pc linear size which are self-opaque in the 21~cm
transition. These objects have \ion{H}{1} column densities as high as
10$^{23}$\cm~ and contribute to a global opacity correction factor of
1.34$\pm$0.05 that applies to the integrated 21~cm emission to obtain a
total \ion{H}{1} mass estimate. High resolution, opacity-corrected
images of the nearest external galaxies have been used to form a
robust redshift zero distribution function of \ion{H}{1},
$f(N_{HI},X,z=0)$, the probability of encountering a specific
\ion{H}{1} column density along random lines-of-sight per unit
comoving distance. This is contrasted with previously published
determinations of $f(N_{HI},X)$ at z~=~1 and 3. A systematic decline
of moderate column density (18 $<$ log(N$_{HI}$) $<$ 21) \ion{H}{1} is
observed with decreasing redshift that corresponds to a decline in
surface area of such gas by a factor of five since z~=~3. The number
of equivalent Damped Lyman Alpha absorbers (log(N$_{HI}$) $>$ 20.3)
has also declined systematically over this redshift interval by a
similar amount, while the cosmological mass density in such systems
has declined by only a factor of two to its current, opacity corrected
value of $\Omega_{HI}^{DLA}$(z=0)~=~5.4$\pm$0.9$\times 10^{-4}$.

We utilise the tight, but strongly non-linear dependance of 21~cm
absorption opacity on column density at z~=~0 to transform our high
resolution \ion{H}{1} images into ones of 21~cm absorption
opacity. These images are used to calculate distribution and
pathlength functions of integrated 21~cm opacity. We suggest that
that this z~=~0 calibration may also apply at higher redshift. In this
case, the incidence of deep 21~cm absorption systems is predicted to
show very little evolution with redshift, while that of faint
absorbers should decline by a factor of five between z~=~3 and the
present. We explicitly consider the effects of \ion{H}{1} absorption
against background sources that are extended relative to the 100~pc
intervening absorber size scale. Extended background sources result in
dramatically altered distribution and pathlength functions which are
insensitive to the predicted redshift evolution. Future surveys of
21~cm absorption will require very high angular resolution, of about
15~mas, for their unambiguous interpretation.

\end{abstract}





\section{Introduction}
The 21~cm neutral hydrogen emission line has proven an extremely useful
diagnostic of the interstellar medium of our own and nearby galaxies
since its first observation by \citet{ewen51} and \citet{mulle51}. And
despite the fact that it was soon recognised \citep{hage54} that this
transition could have significant opacity under interstellar
conditions, it has often proven convenient to neglect that possibility
when estimating the total \ion{H}{1} mass of galaxies or the
statistical properties of the neutral interstellar medium. Opacity
estimation is straightforward in directions toward bright, compact
background continuum sources, but is considerably more challenging to
determine when based on the emission spectrum alone \citep{rohl72}. A
finite line opacity will cause systematic changes to the shape of
emission spectra that can be used to estimate opacity for simple media
and viewing geometries. Line-of-sight confusion encountered when
viewing edge-on systems presents a very serious challenge to such an
analysis given the likelihood for discrete features to overlap in both
position and velocity. This has precluded detailed opacity modelling
of the Galactic neutral hydrogen emission near and in the plane. Only by
assuming a single universal effective temperature for all the
\ion{H}{1} in the Galaxy have some crude opacity corrections been made
(e.g. \citet{schm57}, \citet{hend82}).

The first detailed opacity modelling of the \ion{H}{1} emission in an
external galaxy has recently been undertaken by \citet{brau09} within
M31. This was enabled by achieving 100~pc linear and
2~\kms\ spectral resolution at sub-Kelvin brightness sensitivity. High
quality fits to individual spectra based on a physical model were
possible over the majority of the disk, with the exception of those
regions where the ``warping'' phenomenon of the outer disk leads to
line-of-sight confusion. That study determined that opacity
corrections could locally exceed an order of magnitude in the implied
\ion{H}{1} column density and that these resulted in a 30\% increase
in the global \ion{H}{1} mass estimate. But how representative are
these values and to what extent are they influenced by the relatively
high mean inclination, $i~=~78^\circ$, of the M31 disk or the finite
spatial resolution?

In this paper we extend the analysis of \citet{brau09} to the two
other external galaxies for which the requisite physical resolution
and sensitivity in the 21~cm \ion{H}{1} emission line are available,
M33 and the LMC. These galaxies have the virtue of a much more
favorable disk orientation, with inclinations of about $50^\circ$ and
$25^\circ$ respectively. The extreme proximity of the LMC also permits
linear scales as small as 15~pc to be probed with good
sensitivity. Although small, this sample permits some general
conclusions to be drawn regarding the statistical occurrence of opaque
21~cm \ion{H}{1} emission and \ion{H}{1} column densities.

We begin with a brief description of the observations and their
reduction in \S\ref{sec:obse} and continue with their opacity analysis
in \S\ref{sec:anis}. In \S\ref{sec:disc} we present a statistical
analysis of the occurrence of opacity corrected \ion{H}{1} column
density within all three galaxies and put these in a cosmological
context by considering the prospects for detection of intervening 21~cm
\ion{H}{1} absorbers toward distant background sources. We assume a
distance to \object{M31} of 785~kpc \citep{mcco05}, to \object{M33} of 794~kpc
\citep{mcco04} and to the \object{LMC} of 51~kpc \citep{koer09} throughout. We
further adopt a flat cosmological model with $\Omega_\lambda$~=~0.73 and a
Hubble constant of 71~km~s$^{-1}$Mpc$^{-1}$ \citep{hins09}.

%__________________________________________________________________

\section{Observations}
\label{sec:obse}

\subsection{M31}
The observations and data reduction methods relevant to M31 are
described in detail in \citet{brau09}. The most relevant aspects are
the high quality total power data obtained with the Green Bank
Telescope (GBT) as described by \citet{thil04} combined with a 163
pointing mosaic obtained with the Westerbork Synthesis Radio Telescope
(WSRT) in its ``maxi-short'' configuration. The data cube used for the
spectral modeling analysis had angular resolution of
30\arcsec~(114~pc), velocity resolution of 2.27~\kms\ and RMS noise
level of 1.0 K per channel. The relevant data attributes are summarized in
Table~\ref{tab:parms}.

\subsection{M33}
Total power observations of a $5^\circ\times5^\circ$ region centered
on M33 were carried out with the GBT during 2002 October using the
same setup described by \citet{thil04} for the observations of
M31. Interferometric data was acquired with the Very Large Array (VLA)
in 1997, 1998 and 2001 under the proposal IDs AT206 and AT268. This
consisted of a six pointing mosaic in the B and CS configurations and
an additional 20 pointing mosaic in the D configuration. Some early
results based on the B and CS configuration data were presented in
\citet{thil02} while an independent reduction of all three VLA
configurations has been presented by \citet{grat10}. Joint
deconvolution of all of the VLA pointing data was carried out after
inclusion of the appropriately scaled GBT total power images following
the method employed for the reduction of the M31 data described in
detail by \citet{brau09}. The data cube used for
the spectral modeling analysis had angular resolution of 20\arcsec~(77~pc),
velocity resolution of 1.42~\kms\ and RMS noise level of 2.1~K per channel.

\subsection{LMC}
The observations and data reduction methods relevant to LMC are
described in detail in \citet{kim03}. The most relevant aspects are
the total power data obtained with the Parkes
Telescope and its Multi-Beam 21~cm receiver combined with a 1344
pointing mosaic obtained with the Australia Telescope Compact Array
(ATCA) in its 750A, 750C and 750D configurations. The data cube used for
the spectral modeling analysis had angular resolution of 60\arcsec~(15~pc), 
velocity resolution of 1.65~\kms\ and RMS noise level of 2.5~K per channel.

\begin{deluxetable}{cccccc}
\tablewidth{0pt}
\tablecaption{Galaxy sample for spectral analysis}
\tablehead{
\colhead{Name}           & \colhead{Beam (arcsec)}      &
\colhead{Beam (pc)}   & \colhead{$\Delta T_B$ (K)}    & \colhead{$\Delta V$ (\kms)} & \colhead{Ref} }      
\startdata
M31 & 30 & 114 & 2.27 & 1.0 & 1 \\
M33 & 20 & 77 & 1.42 & 2.1 & 2 \\
LMC & 60 & 15 & 1.65 & 2.5 & 3 \\
\enddata
\label{tab:parms}
\tablerefs{
(1) Braun et al. 2009; (2) Thilker et al. 2002; (3) Kim et al. 2003}
\end{deluxetable}

\subsection{Additional Targets}
\label{sec:addt}
While it would be very advantageous to extend our galaxy sample beyond
the Local Group targets listed above, this is beyond the capabilities
of current surveys and radio telescopes. 

For comparison, the THINGS survey \citep{walt08} consists of 
\ion{H}{1} imaging of 34 relatively nearby galaxies using multiple
configurations of the VLA. Although this survey achieves an angular
resolution as high as 6~arcsec, which corresponds to less than 150~pc
for the thirteen THINGS galaxies within about 4~Mpc, the corresponding
RMS brightness sensitivity per velocity channel varies between about
10 and 20~K at this angular resolution while the velocity resolution
is too coarse for several of the galaxies (5.2~\kms) and only marginal
(2.6~\kms) for the remainder. As will become apparent below,
parameterised spectral fitting is only found to be viable when at
least one independent velocity channel exceeds 10$\sigma$ and at least
five exceed 5$\sigma$. For this reason it is vital to achieve better
than about 2~K brightness sensitivity over 2~\kms\ at the required
physical resolution of about 100~pc that matches intrinsic \ion{H}{1}
structure sizes. The THINGS survey has insufficient sensitivity by a
factor of between 5 and 10 for this purpose. Remedying this
shortcoming would require between 25 and 100 times longer
integrations, or between 250 and 1000 hours per target with the EVLA.

Another ongoing survey of nearby galaxies that might be considered of
potential relevance is the HALOGAS survey of \citet{heal11}. That
survey involves relatively deep (120 hours per target) observations of
a galaxy sample within about 10~Mpc. Unfortunately, as that survey is
designed primarily to detect diffuse gas in the extended environments
of galaxies it has only modest physical (about 1~kpc) and velocity
(4.2~\kms) resolution that make it unsuitable for parameterised
spectral fitting of well-resolved structures.

\section{Spectral Analysis}
\label{sec:anis}
The method of spectral analysis has been described in detail in
\citet{brau09}. We repeat the most relevant aspects of that analysis
below for clarity. 

Each line profile is modeled as a spatially resolved {\it
  isothermal\ } \ion{H}{1} feature in the {\it presence of turbulent
  broadening\ } as
\begin{equation}
T_B(V) = T_S\{1-exp[-\tau(V)]\}
\label{eqn:tb}
\end{equation}
with
\begin{equation}
\tau(V) = {5.49\times10^{-19}N_{HI} \over
  T_S\sqrt{2\pi\sigma^2}} exp\Biggl(-0.5{V^2 \over
  \sigma^2}\Biggr)
\label{eqn:tau}
\end{equation}
where the velocity dispersion, $\sigma$, has units of \kms\ and is the
quadratic sum of an assumed thermal and nonthermal contribution
$\sigma^2 = \big(\sigma_T^2+\sigma_{NT}^2\big)$ with the thermal
contribution given by $\sigma_T=0.093 \sqrt T_k$, for a kinetic
temperature, $T_k$. Such profiles have a Gaussian shape for low
ratios of the \ion{H}{1} column density (in units of cm$^{-2}$) to
spin temperature, $N_{HI}/T_S$, but become increasingly flat-topped
when this ratio becomes high. 

We found the best-fitting model spectrum for the brightest spectral
feature along each line-of-sight when the observed peak brightness
temperature exceeded 10 times the RMS noise and the truncated peak
(where brightnesses greater than 50\% of the peak were encountered)
spanned at least five independent velocity channels. The requirement
for a well-sampled linewidth translates directly into one on the
velocity resolution of the input galaxy data of better than about
2~\kms. The truncation was done to both isolate single spectral
components from possibly blended features as well as eliminating
potential broad wings from the fit. In this way, the broad wings of
secondary components along the same line of sight, for example from
additional warm neutral gas with $\sim10^4$~K temperature, could be
excluded from the fit. The data were compared to a pre-calculated set
of model spectra spanning the range log(N$_{HI}$) = 20.0 to 23.5 by
0.01, log(T$_C$) = 1.2 to 3.2 by 0.01 and log($\sigma_{NT}$) = 0.3 to
1.5 by 0.04. A search in velocity offset was done over displacements
of $-$4 to +4 by 1~\kms\ with respect to the line centroid as
estimated by the first moment of each truncated peak.

Only those fits with a normalized $\chi^2 = \Sigma_i
(D_i-M_i)^2/(N\sigma^2) < 25$ were retained; a cut-off based on the
rejection level needed within M31 to exclude instances of line
blending where the outer ``warp'' is seen in projection against the
inner disk. In the case of M31 this resulted in rejection of some 4\%
of the fits, while for both M33 and the LMC the fit rejection rate
dropped to below 1\%.

Results of the spectral fitting were recorded as images of the
physical parameters, $N_{HI}^{Fit}, T_S$ and $\sigma_{NT}$ together
with the integral of brightness temperature over velocity that
corresponds to the column density of the fit, $\int T_B^{Fit}dV$. This
allowed calculation of a total corrected column density image from,
\begin{equation}
N_{HI}^{Tot} = N_{HI}^{Fit} +
1.823\times10^{18}\biggl(\int T_B^{Tot}dV-\int T_B^{Fit}dV\biggr)
\end{equation}
where $\int T_B^{Tot}dV$ is just the usual image of integrated
emission. Those lines-of-sight with insufficient fit quality ($\chi^2
> 25$) or insufficient peak brightness and width (at least one
independent velocity channel exceeding 10$\sigma$ and at least five
exceeding 5$\sigma$) were assumed to have negligible opacity and
simply assigned the optically thin value. 

   \begin{figure*}
   \centering
   \includegraphics[width=16.5cm]{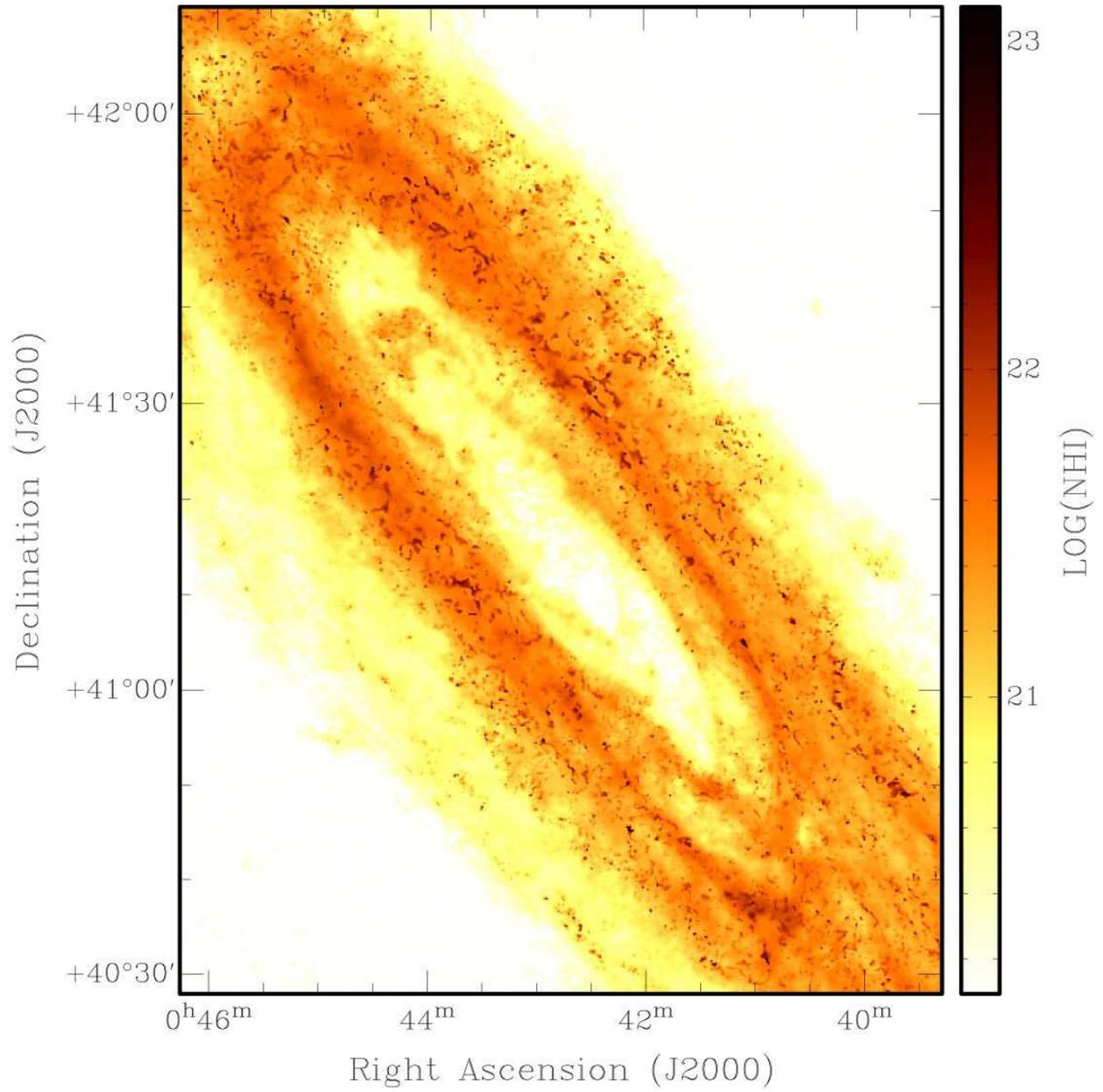}
      \caption{Opacity corrected neutral hydrogen column density in
        the inner disk of M31. }
         \label{fig:m31nhc}
   \end{figure*}

   \begin{figure*}
   \centering
   \includegraphics[width=16.5cm]{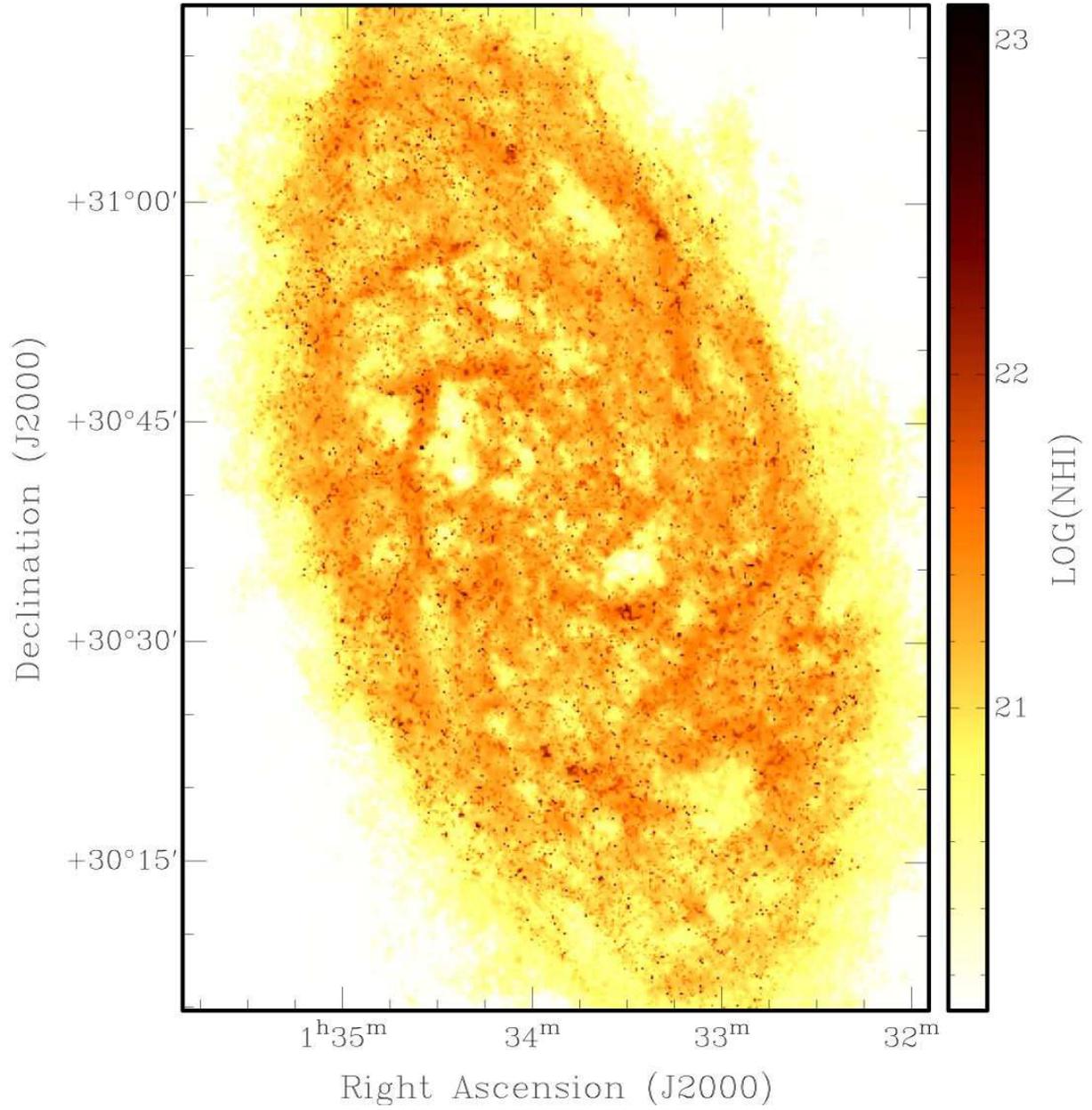}
      \caption{Opacity corrected neutral hydrogen column density in M33.}
         \label{fig:m33nhc}
   \end{figure*}

   \begin{figure*}
   \centering
   \includegraphics[width=16.5cm]{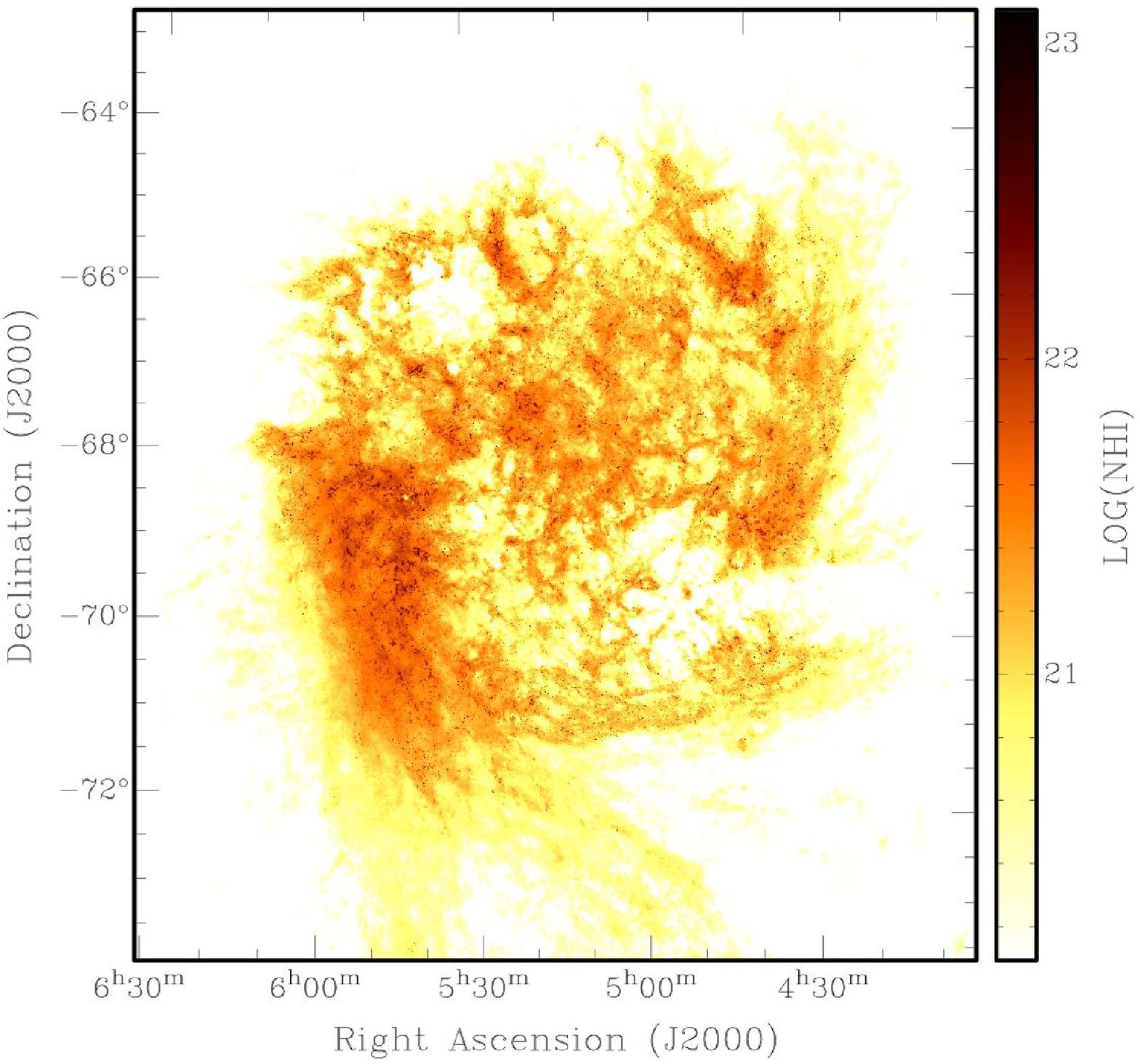}
      \caption{Opacity corrected neutral hydrogen column density in the LMC.}
         \label{fig:lmcnhc}
   \end{figure*}

\section{Discussion}
\label{sec:disc}

\subsection{Opacity-corrected images and the local $\Omega_{HI}^{gal}$}

\begin{figure*}
   \centering
   \includegraphics[trim=0 0 32 0,clip=true,width=8cm]{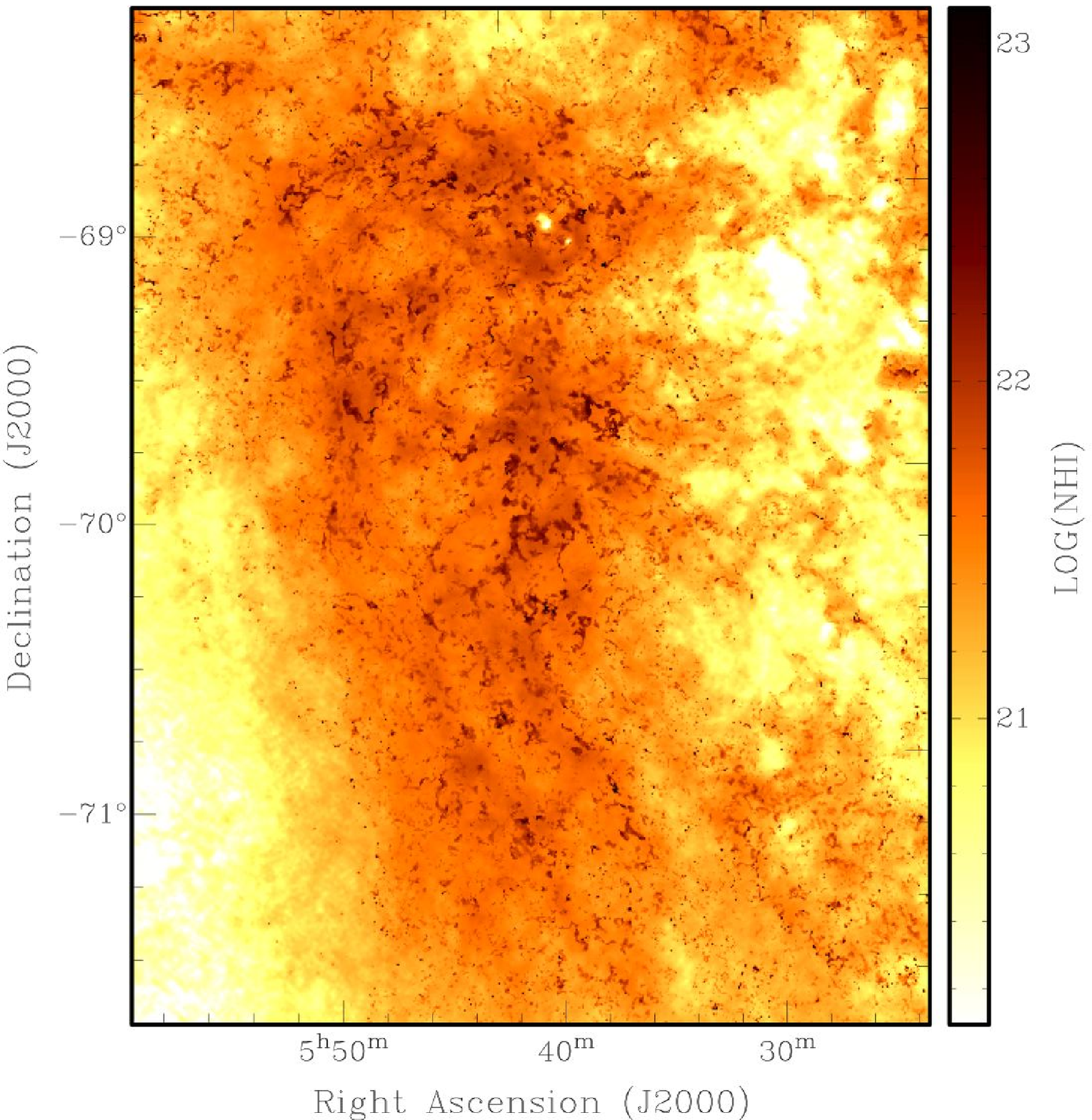}
   \hfill
   \includegraphics[trim=32 0 0 0,clip=true,width=8cm]{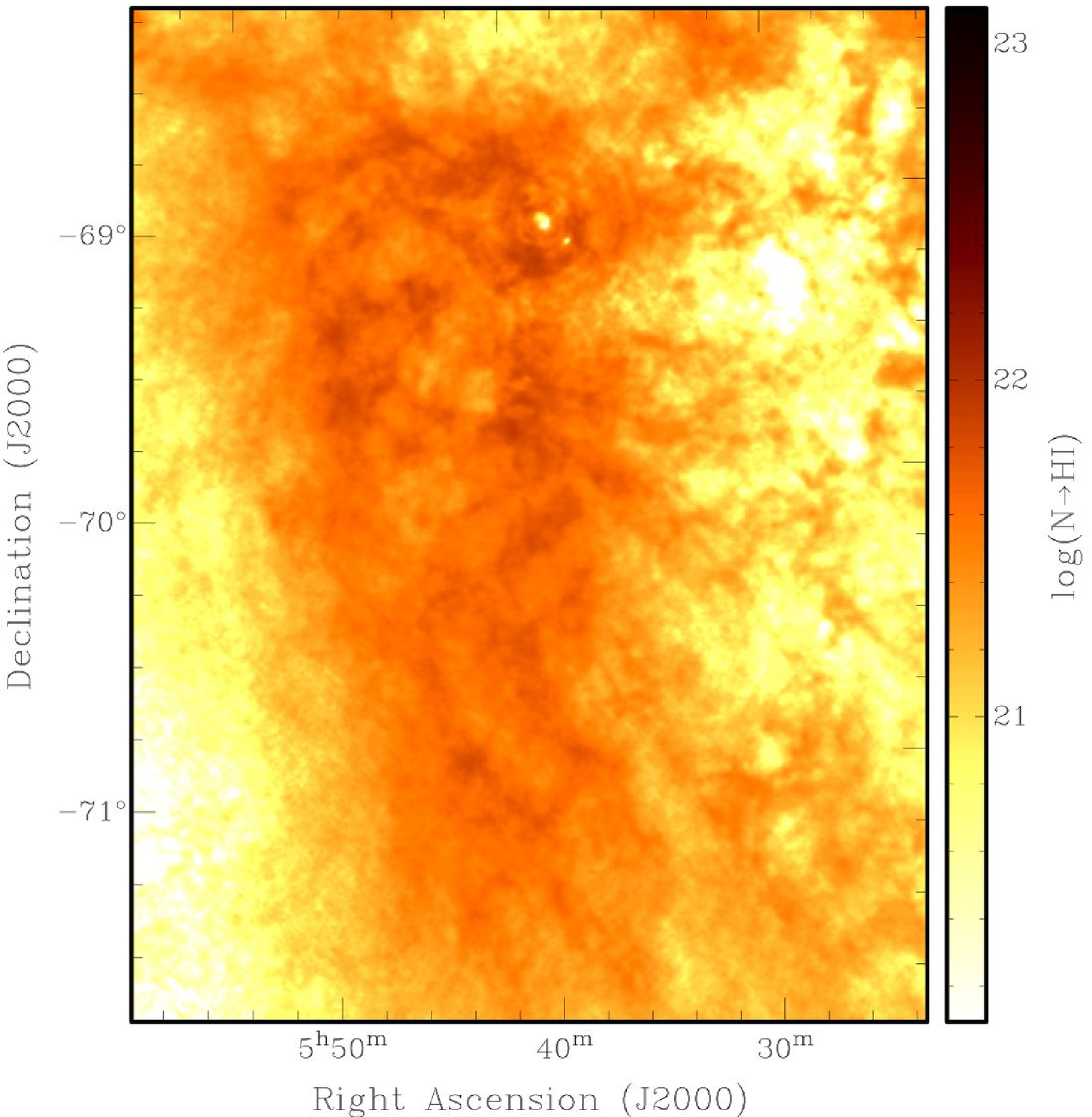}
   \caption{Opacity corrected (left) and apparent (right) neutral
     hydrogen column density in the southeastern quadrant of the LMC.}
   \label{fig:lmcnhcz}
   \end{figure*}

We present images of the opacity corrected neutral hydrogen column
density in Figures~\ref{fig:m31nhc}--\ref{fig:lmcnhc}. In contrast to
images of the apparent column density, $N'_{HI} = 1.823\times10^{18}
\int T_BdV$, which saturate below 10$^{22}$~cm$^{-2}$ into an almost
featureless fog, there is widespread fine structure that extends to
about 10$^{23}$~cm$^{-2}$. It is interesting to note that comparably
high \ion{H}{1} columns have been observed in the Lyman Alpha
absorption spectra of high redshift GRBs \citep{fynb09}.  This
contrast is illustrated in Figure~\ref{fig:lmcnhcz} where the corrected
and apparent distributions are shown side by side for a region in the
southeastern quadrant of the LMC that includes 30 Doradus (near
($\alpha,\delta$)=(05:40,$-$69:00)). This figure also serves to
demonstrate the exceptional detail provided by the LMC linear
resolution of only 15~pc. While many of the high column density
features in M31 and M33 are only marginally resolved with 100~pc
resolution, the majority of the LMC features are well-resolved into
structures that have coherent derived properties, despite the fact
that every line-of-sight has been independently fit with a model
spectrum. The corrected total \ion{H}{1} masses exceed those that
follow from the assumption of negligible 21~cm opacity by factors of
1.30, 1.36 and 1.33 for M31, M33 and the LMC respectively. As noted by
\citet{brau09}, the M31 value may be biased to a lower value since
some portions of the disk are confused by the warp and could not be
adequately modeled. Although our sample of galaxies with
opacity-corrected \ion{H}{1} column densities is very small, it does
sample both the mass density peak and the highly populated tail of the
\ion{H}{1} mass function \citep{zwaa03} with apparent values of
$log(M_{HI})$~=~9.77, 9.13 and 8.61. And while all three galaxies are
of moderately late type, the same is true of about 90\% of galaxies
within the complete HIPASS sample \citep{zwaa03} both by their
contribution to the total \ion{H}{1} mass and in terms of space
density.  If these global correction factors are representative of
galaxies in general, they suggest that the estimated redshift zero
mass density, $\Omega_{HI}^{gal}$, determined from unbiased surveys of
\ion{H}{1} emission from galaxies in the local universe requires this
magnitude of upward correction. Taking the average of the HIPASS
determined value \citep{zwaa03} and the recent ALFALFA value
\citep{mart10} after converting both to the assumed cosmology and
scaling up by an opacity correction factor of 1.34$\pm$0.05 yields
$\Omega_{HI}^{gal}(z=0)$~=~5.6$\pm0.9 \times 10^{-4}$.

The existence of additional gas, in excess of that inferred from the
published 21~cm and CO data has been previously suggested by
\citet{bern08} in the case of the LMC from what has been termed
``excess FIR'' emission. Those authors have demonstrated that the dust
optical depth is enhanced in regions of the highest apparent
\ion{H}{1} column density consistent with approximately twice the
total gas mass being present. The opaque 21~cm emission that we
document appears to account for some 35\% of this FIR excess. This
cool atomic hydrogen gas will likely be closely associated with
diffuse molecular hydrogen gas that may be deficient in CO emission
relative to the high density, self-gravitating molecular clouds that
are generally used to calibrate the CO to H$_2$ conversion factor
\citep[e.g.][]{dick86}. A detailed analysis of atomic, molecular and
dust tracers in the LMC, including our opacity-corrected \ion{H}{1}
results will be presented in \citet{thil12}.

\subsection{\ion{H}{1} distribution and mass density functions}

\begin{figure*}
   \centering 
   \includegraphics[trim=0 0 50 0,width=8.cm]{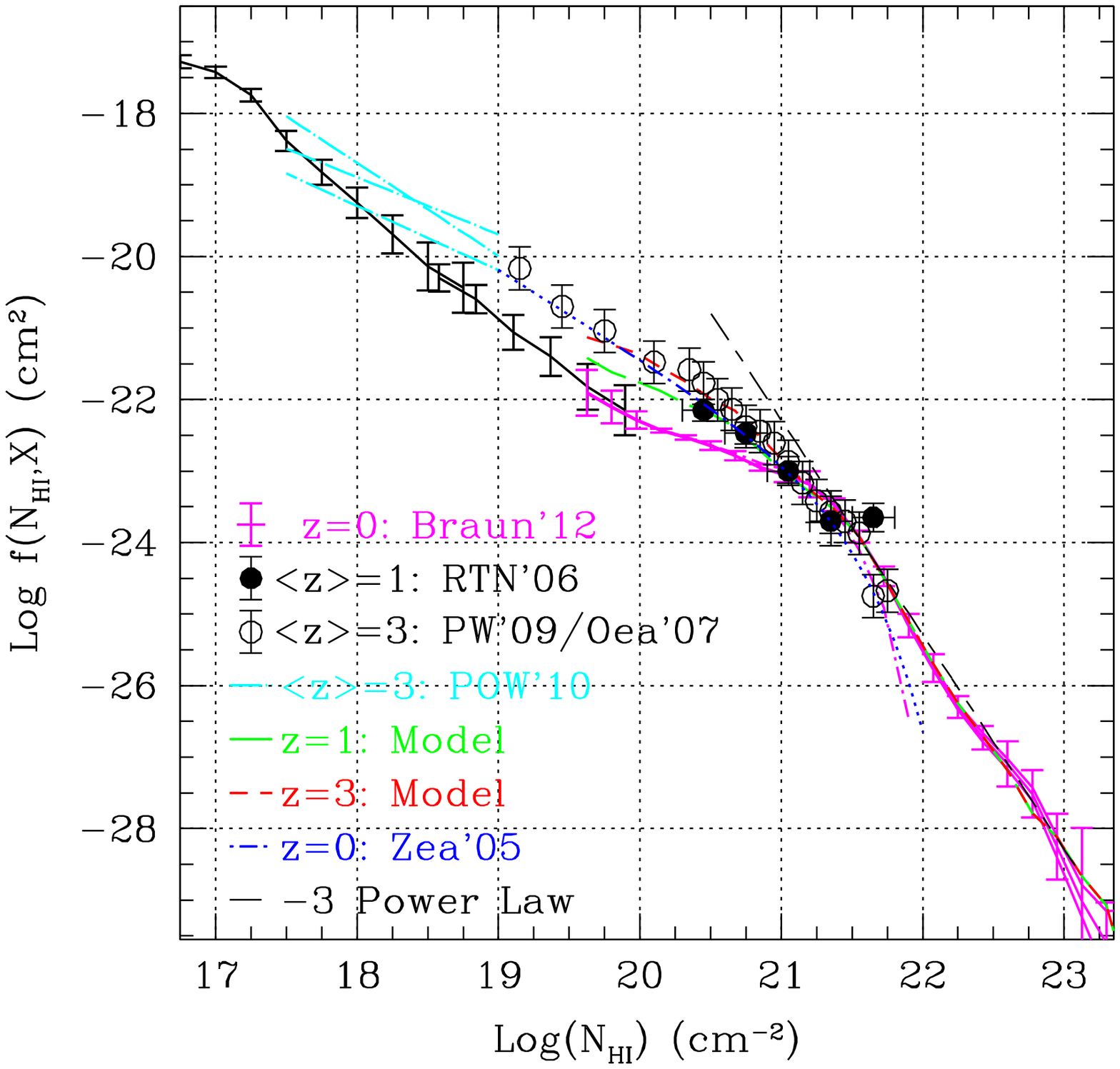}
   \hfill
   \includegraphics[trim=50 0 0 0,width=8.cm]{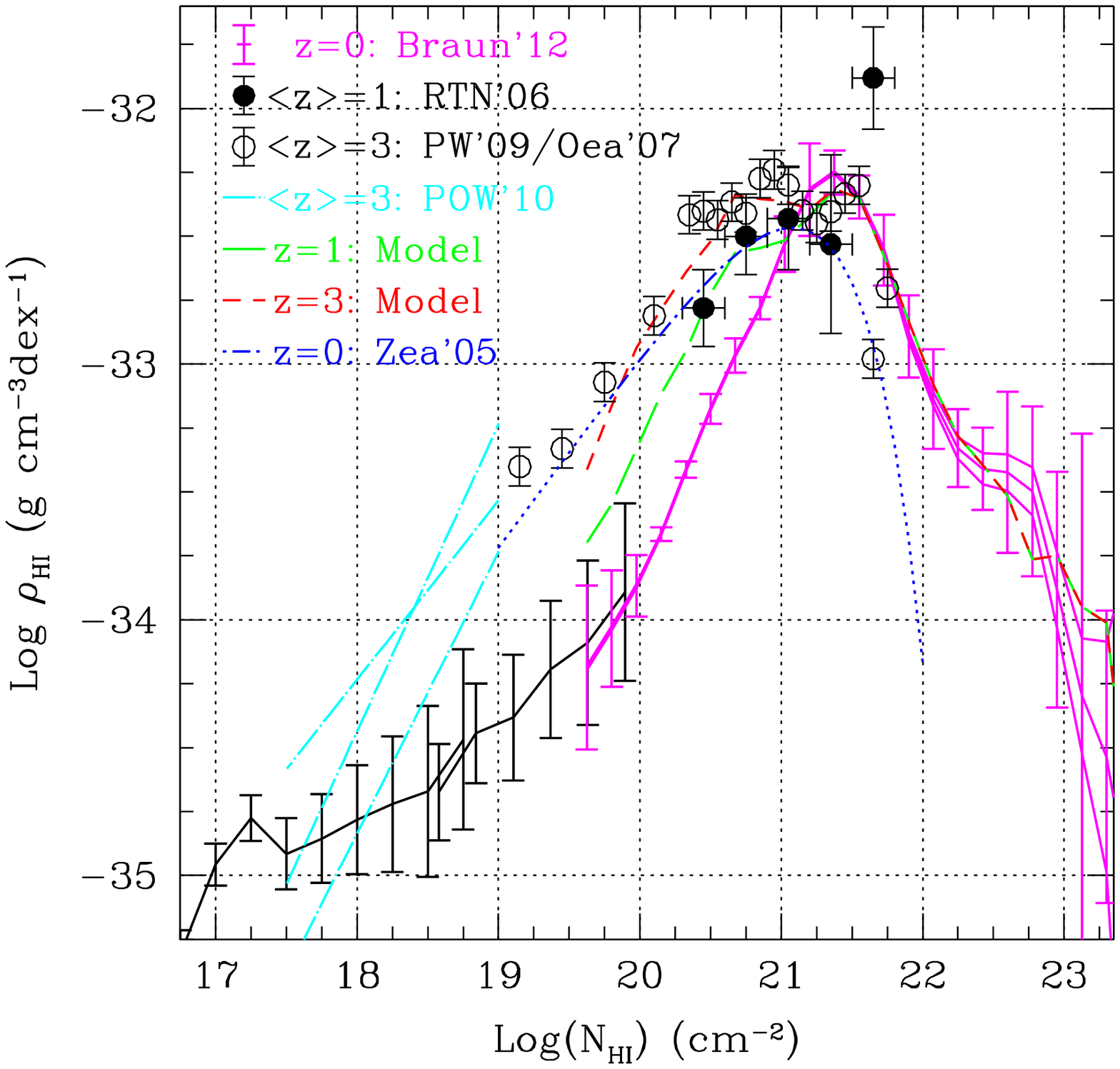}  
   \caption{Distribution function of neutral hydrogen column density
     (left) and the mass density function (right). Please see
     Figure~\ref{fig:fnhcz} for greater detail of the inner portion of
     this figure. Solid curves represent our current, z~=~0,
     determination. Filled circles are used for the $<z>~=~1$ QSO
     absorption line data of \citet{rao06} and open circles for the
     $<z>~=~3$ QSO absorption line data of \citet{proc09} and
     \citet{omea07}. The dot-long-dash curves extend the $<z>~=~3$
     data to lower $N_{HI}$ \citep{proc10}. The dot-dash curve is the
     z~=~0 determination of \citet{zwaa05} based on low resolution
     (1.4~kpc) \ion{H}{1} emission images. The long-dashed and dashed
     curves represent the model we develop to represent the $<z>~=~1$
     and 3 distributions. The short-dash, long-dash curve in the
     left panel is a $-3$ power law overlaid on the high $N_{HI}$ data.}
         \label{fig:fnhc}
   \end{figure*}

It has been recognised for some time that images of 21~cm \ion{H}{1}
emission from nearby galaxies can be used to construct a distribution
function that documents the probability to intercept a particular
column density on a random line-of-sight. The first application of
this approach \citep{rao93} employed the assumption that the apparent
\ion{H}{1} mass was distributed smoothly over a typical \ion{H}{1}
size that depends on galaxy type. Subsequent studies
(e.g. \citet{ryan03}, \citet{zwaa05}) have made use of interferometric
images to estimate the areal coverage of specific column densities,
but even these have been limited to a median physical resolution of
some 1.4~kpc and no ability to correct for 21~cm opacity effects. Most
recently, \citet{erka12} have used the THINGS galaxy images to this end,
despite the fact that those data have only marginal resolution and
inadequate sensitivity to enable resolved opacity
corrections to be made, as discussed in \S\ref{sec:addt} above. We
have used the opacity corrected column density images
(Figs.\ref{fig:m31nhc}--\ref{fig:lmcnhc}) together with the sensitive
wide-field total power 21~cm \ion{H}{1} emission data from the combined
M31/M33 system \citep{brau04} and deep GBT imaging of the M31
environment \citep{thil04} to construct the \ion{H}{1} distribution
function shown in Figure~\ref{fig:fnhc} that spans more than six orders
of magnitude in \ion{H}{1} column density from 10$^{17}$ --
10$^{23}$cm$^{-2}$. 

   \begin{figure*}
   \centering 
   \includegraphics[trim=0 0 50 0,width=8.cm]{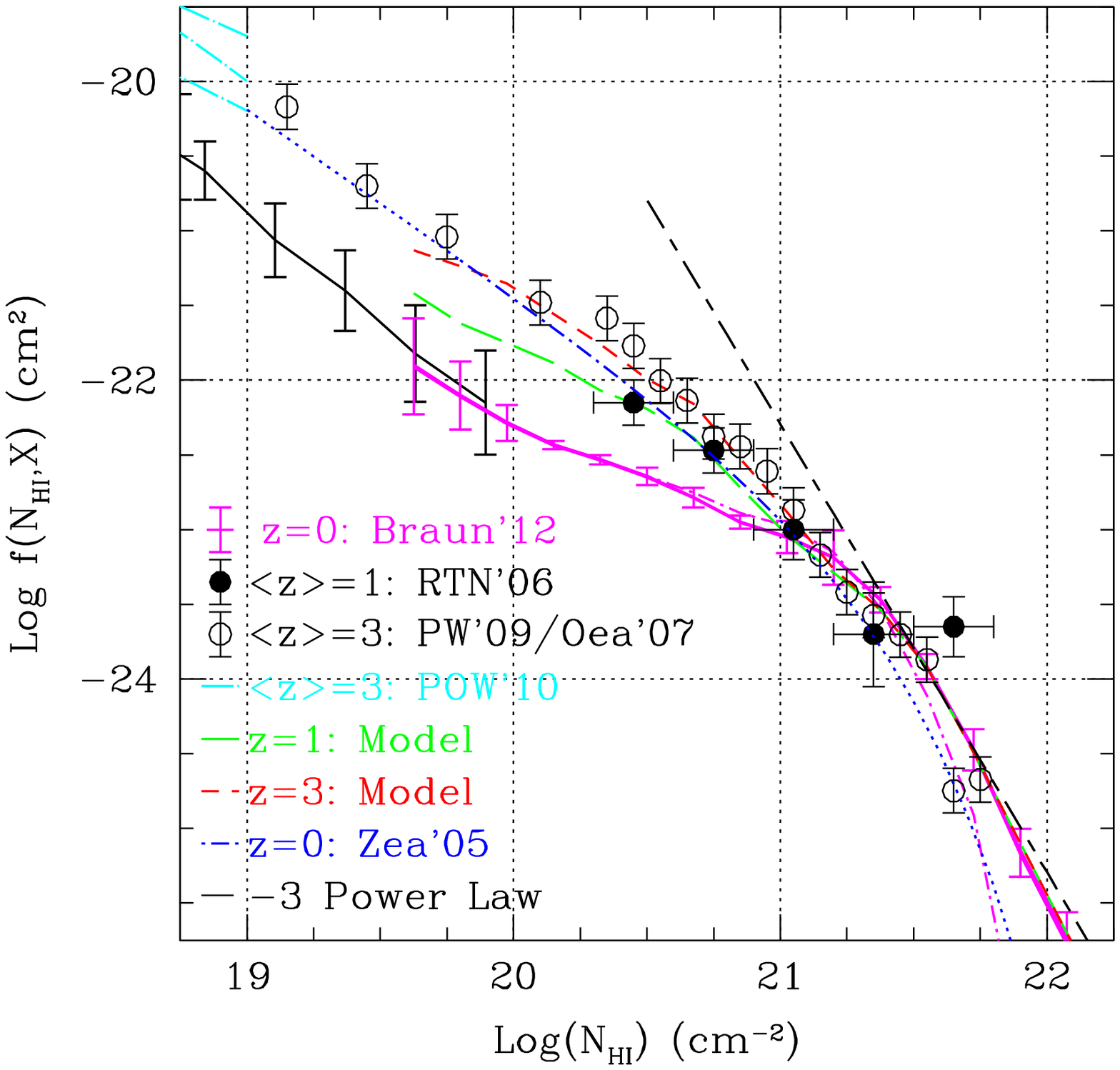}
   \hfill
   \includegraphics[trim=50 0 0 0,width=8.cm]{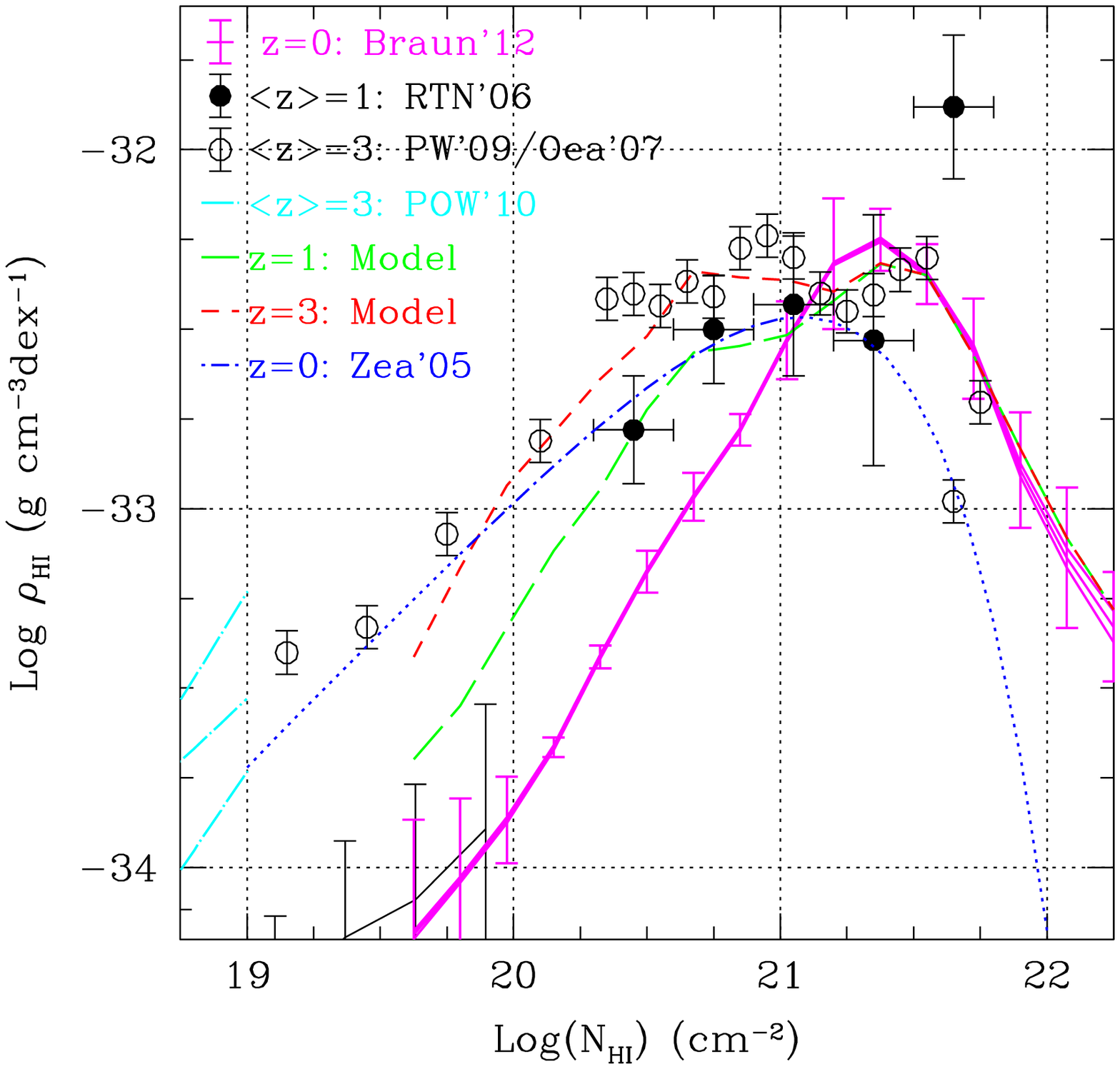}  
   \caption{Distribution function of neutral hydrogen column density
     (left) and the mass density function (right) as in
     Figure~\ref{fig:fnhc}, but limited to a smaller range in \ion{H}{1}
     column density. }
         \label{fig:fnhcz}
   \end{figure*}

The curves in  Figure~\ref{fig:fnhc} were calculated from,
\begin{equation}
f(N_{\rm HI},X) = {c \over H_0}{\rho_{av} \over \Sigma_i\rho_{i}} \Sigma_i \theta_{i}
{A({\rm log}(N_{\rm HIi})) \over N_{\rm HI}{\rm ln}(10) d {\rm log}(N_{\rm HI})}\ \ \  {\rm cm}^2
\label{eqn:fnhc}
\end{equation}
where X represents unit co-moving depth along the line-of-sight,
\begin{equation}
dX = {H_0 \over H(z) } (1+z)^2 dz
\label{eqn:dXdz}
\end{equation}
and
\begin{equation}
\theta_{i} = \theta_* {\rm ln}(10) ({\rm M_{i}}/ {\rm
M_*})^{\alpha+1} {\rm exp}(-{\rm M_{i}/M_*})
\label{eqn:tmi}
\end{equation}
is the space density of galaxies with an {\it apparent}
(i.e. uncorrected for opacity effects) \ion{H}{1} mass $\rm M_{i}$ in
units of Mpc$^{-3}$dex$^{-1}$ and $A({\rm log}(N_{\rm HIi}))$ is the
surface area subtended by \ion{H}{1} in the column density interval
$d{\rm log}(N_{\rm HI})$ centered on $N_{\rm HI}$ for galaxy $i$. The
area is measured by counting image pixels in each column density range
and multiplying by the physical area of each pixel.  $\rho_{i} =
\theta_{i} {\rm M_{i}}$ is the mass density due to galaxies of type
$i$. The HIMF parameters were taken from \citet{zwaa03}; namely a
faint-end slope, $\alpha=-1.30\pm0.08$, characteristic \ion{H}{1}
mass, log(M$_*/$M$_\odot)=9.84\pm0.06$ and normalization,
$\theta_*=7.3\pm1.8\times10^{-3}$ Mpc$^{-3}$dex$^{-1}$, where $H_0=71$
km$^{-1}$Mpc$^{-1}$ has been assumed throughout \citep{hins09} and
been used to rescale the HIMF parameters. The term, $\rho_{av} /
\Sigma_i\rho_{i}$ in eqn.~\ref{eqn:fnhc} is introduced to normalize
our measurements for three galaxies, to the global average density of
{\it apparent} \ion{H}{1} in galaxies, $\rho_{av}=
5.8\times\pm1.0\times10^7$M$_\odot$Mpc$^{-3}$, also taken from Zwaan
et al. We stress that we make use of consistent quantities for this
calculation, namely {\it apparent} galaxy \ion{H}{1} masses and mass
densities based on {\it apparent} mass. Alternatively, {\it corrected}
quantities could have been used throughout.

The solid curve in Figure~\ref{fig:fnhc} below log(N$_{HI}$)~=~18.7 is
taken from the wide-field ($60^\circ\times30^\circ$) WSRT total power
data \citep{brau04}, the curve below log(N$_{HI}$)~=~19.8 from the
extended ($7^\circ\times7^\circ$) GBT total power data of M31
\citep{thil04}, while the curve above log(N$_{HI}$)~=~19.5 is that
derived from the opacity corrected high resolution images presented in
Figs.~\ref{fig:m31nhc}--\ref{fig:lmcnhc}. For comparison, we also plot
the distribution function without opacity correction as the dark
dot-dash curve in the figure. The uncorrected data show an exponential
upper cut-off that precludes column densities as large as
10$^{22}$cm$^{-2}$. The opacity corrected data show an approximate
power law extension to the distribution above about
log(N$_{HI}$)~=~21.5. We plot a power law with $-3$ index as the
dot-dash line that provides a very good match to the measured
distribution. Perhaps fortuitously, this power law index corresponds
to what would be expected for a population of resolved thin structures
with some characteristic face-on column density when viewed at random
orientations, as first noted by \citet{milg88}. This may well be a
plausible interpretation for observations such as presented here which
resolve the atomic ``skins'' of what are likely to be randomly
oriented clouds of molecular hydrogen on 100~pc scales within galaxy
disks.

Of particular note is the magnitude of the statistical errors in our
determination of $f(N_{HI},X)$. The solid curve above
log(N$_{HI}$)~=~19.5 in the figure have been plotted three times: once
as measured and again with both the addition and subtraction of the
statistical 1$\sigma$ error calculated from the number of independent
image pixels contributing to each column density bin. Only above
log(N$_{HI}$)~=~22.5 can these curves be distinguished from one
another at all. The detailed shape of the combined curve above
log(N$_{HI}$)~=~19.5 is also reproduced in detail by the three
galaxies individually, despite the fact that they represent different
Hubble types and \ion{H}{1} masses; ranging from SB(s)m, to SA(s)cd to
SA(s)b and spanning an uncorrected $log(M_{HI})$~=~8.6 -- 9.8. From
Figure~13 of \citet{zwaa03} it is apparent that galaxies of these
relatively late types contribute about 90\% to the \ion{H}{1} mass
function in terms of both total mass and galaxy space density.  The
degree of inter-galaxy consistency is demonstrated with the error bars
on the solid curve which represent RMS differences of the distribution
functions of the three individual galaxies with respect to the
combined $f(N_{HI},X)$. What hasn't been included in these error bars
is the systematic error due to the uncertainty in the normalisation
for galaxy space density, $\theta_*$ (and hence $\rho_{HI}$), of about
0.1 dex. Errors in the normalisation would result in a vertical shift
of the entire distribution function, but would not otherwise change
the shape.

We plot the parameterised fit to the \ion{H}{1} distribution function
of \citet{zwaa05} as the short-dashed curve in Figs.~\ref{fig:fnhc}
and \ref{fig:fnhcz}. That curve is based on the analysis of a sample
of 355 galaxy images in 21~cm \ion{H}{1} emission made as part of the
``WHISP'' project \citep{vand01}. The distribution of Hubble types in
the WHISP sample \citep{hold11} is similar to that seen in
the complete HIPASS \citep{zwaa03} sample, consisting of about 90\%
galaxies later than Sb and 10\% earlier. However, as noted at the outset of
this section, it is limited by a median physical resolution of about
1.4~kpc and the inability to account for opacity effects. The authors
suggest that their data might be representative in the range
log(N$_{HI}$)~=~19.8 to 21, where sensitivity limits the low end and
resolution limits the high end of that range. We plot the extrapolated
extension of the fit beyond that range as a dotted line in the
figure. \citet{zwaa05} have explored the consequences of smoothing of
the WHISP data to even coarser resolutions and note that this results
in a steepening of the apparent $f(N_{HI},X)$ for
log(N$_{HI}$)~=~20 -- 21. Conversely, they expect that achieving higher
physical resolution would lead to a flattening of the derived
$f(N_{HI},X)$ in that column density range, which would then more
closely resemble what we observe. It should be noted that the ratio of
median beam area between the WHISP sample and ours is a factor of
about 400, implying that we sample about three times as many
independent beam areas in our sample of only three galaxies as exist
in the entire 355 galaxy WHISP sample. 

\begin{figure*}
     \centering
     \includegraphics[trim=25 0 75 0,height=5.5cm]{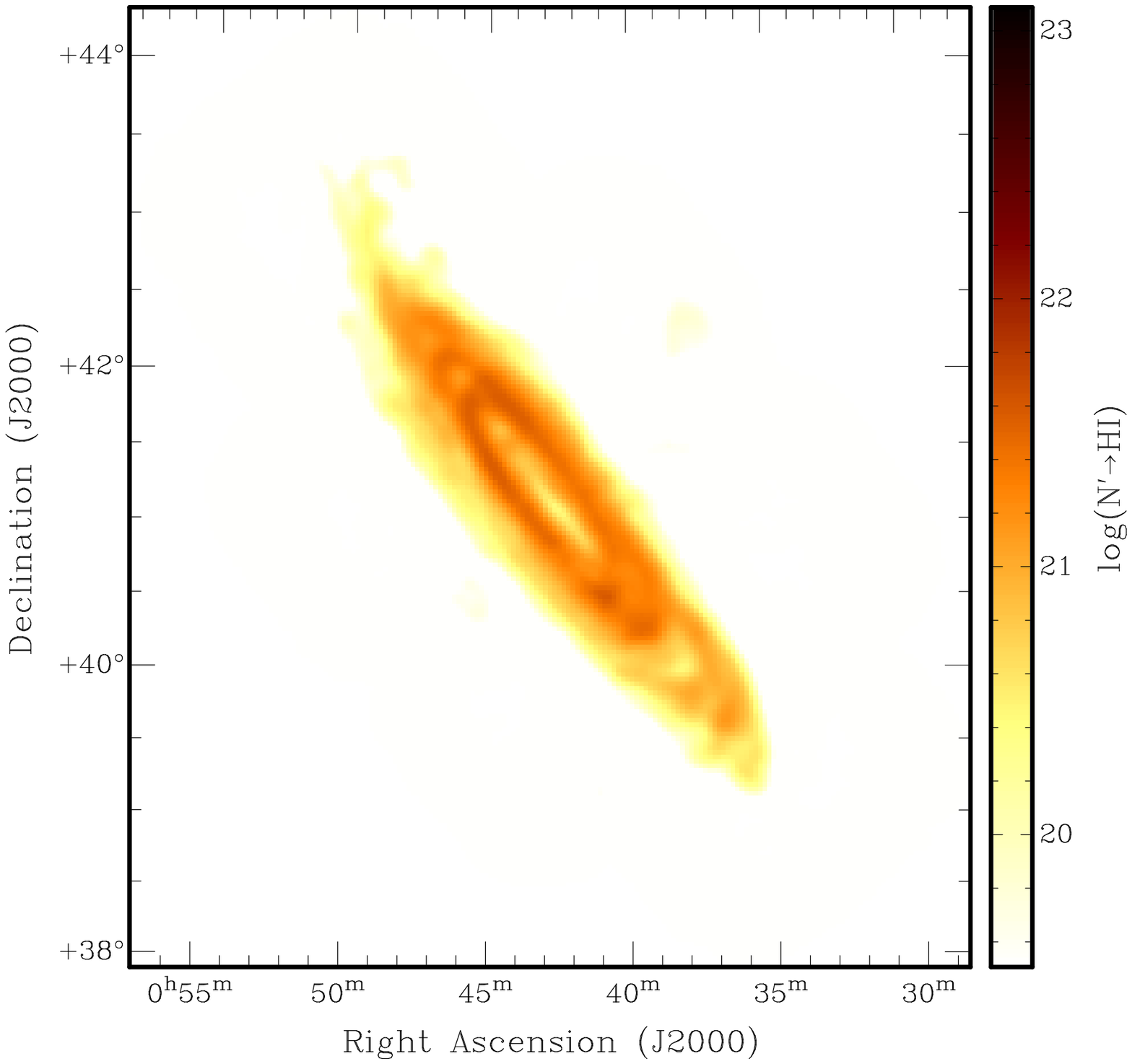}  
     \hfill
     \includegraphics[trim=50 0 50 0,height=5.5cm]{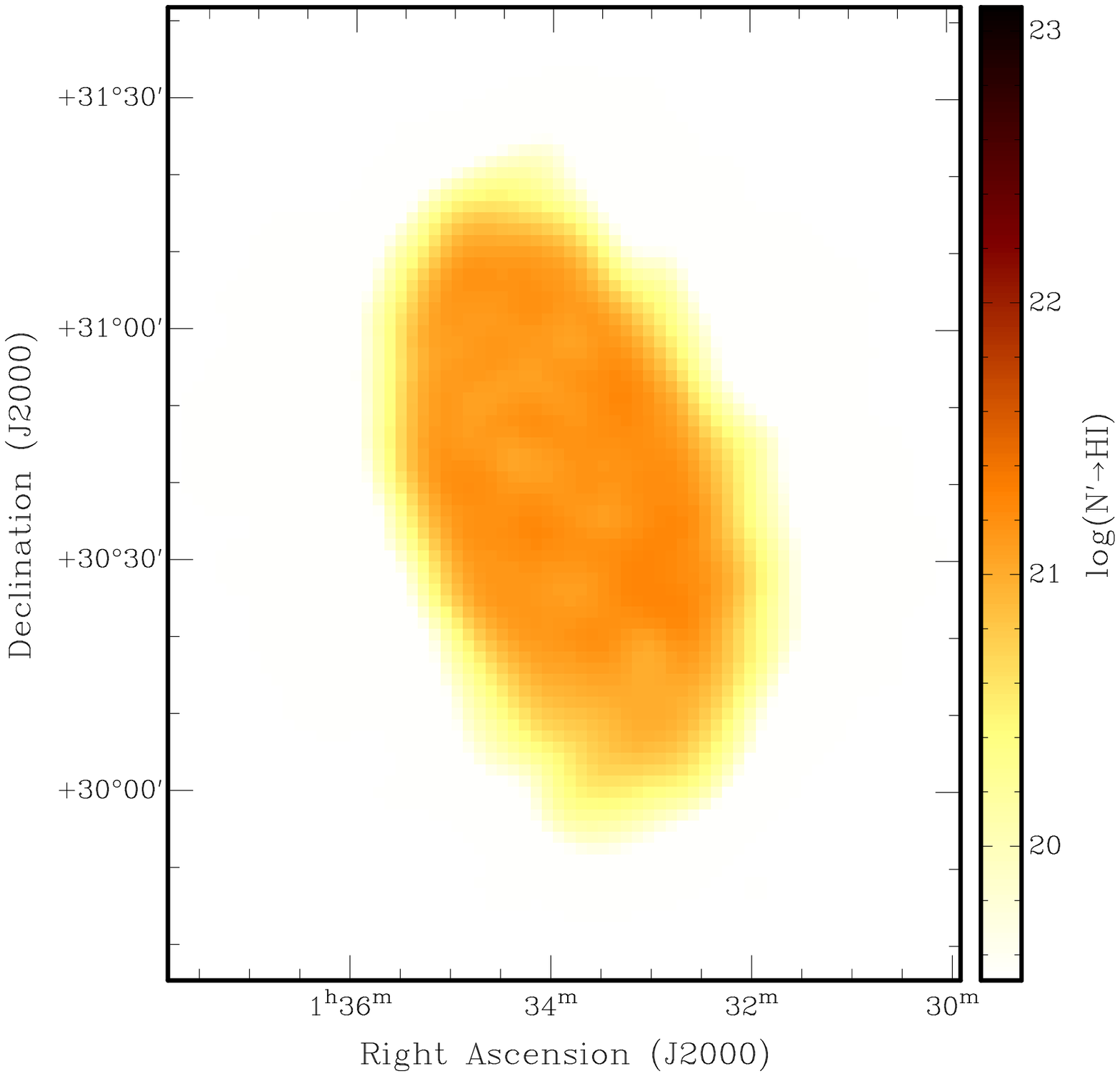}  
     \hfill
     \includegraphics[trim=75 0 25 0,height=5.5cm]{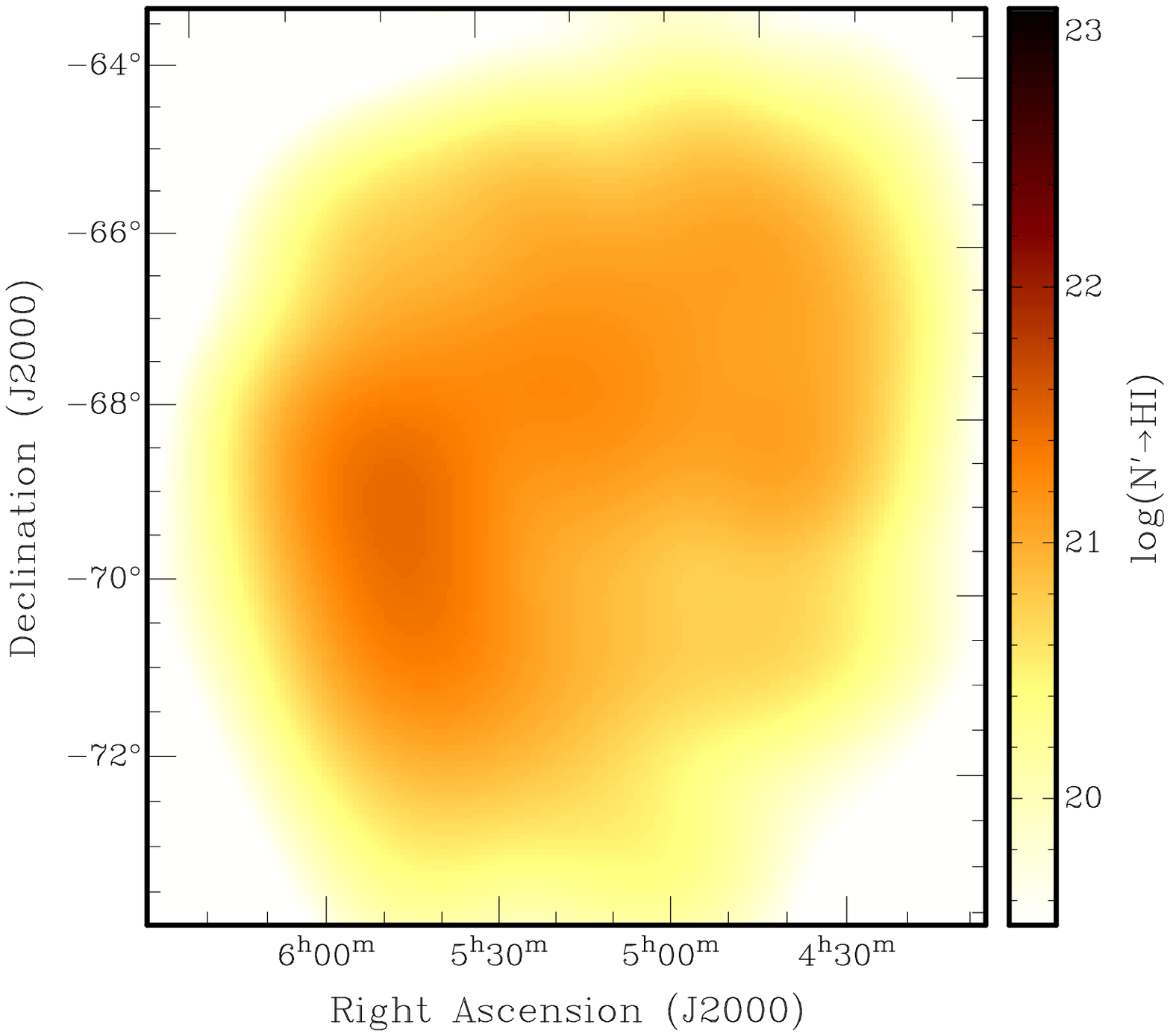}  
     \caption{Apparent \ion{H}{1} column density of M31 (left), M33
       (center) and the LMC (right) as seen with an angular resolution
       that corresponds to 1.4~kpc, the median physical resolution of the
       ``WHISP'' survey used in the \citet{zwaa05} study. }
     \label{fig:Nhw}
   \end{figure*}

We have attempted to reproduce the effects of coarse physical
resolution by smoothing the uncorrected (for opacity) images of our
three galaxies to the median beam size of the WHISP sample, 1.4~kpc,
and display these in Figure~\ref{fig:Nhw}. As is clear from a comparsion
of Figures~\ref{fig:m31nhc}--\ref{fig:lmcnhc} with Figure~\ref{fig:Nhw},
there is little useful information regarding the statistical ocurrence
of intrinsic \ion{H}{1} column densities at such a coarse physical
resolution, given the extensive smearing of intrinsic peaks into
larger areas of apparently lower column density. Although the combined
distribution function based on these three degraded images yields
better agreement with the \citet{zwaa05} curve, it still does not
agree in detail. Looking at the $f(N_{HI},X)$ contributions from the
individual galaxies, it becomes apparent that the consequence of
spatial smearing does not impact all galaxies to the same
degree. While the redistribution of actual to apparent column
densities has a moderate impact for the larger galactic systems, it
becomes more dramatic for physically smaller ones. In the case of the
LMC, the apparent area covered by log(N$_{HI}$)~=~20 -- 21 gas is
increased by about a factor of two by this smoothing. Its likely that
this apparent enhancement of such column densities will be even more
extreme in lower \ion{H}{1} mass systems since they tend to have
even smaller physical size. Since low \ion{H}{1} mass systems strongly
dominate total galaxy numbers, they will similarly influence the
distribution function of a galaxy sample via their relative weighting,
$\theta_i$ in eqn.~\ref{eqn:fnhc}. While we are not able to reproduce
the \citet{zwaa05} results in detail, it seems likely that the
discrepancies are consistent with insufficient physical resolution in
the Zwaan et al. study. 

\citet{erka12} have presented \ion{H}{1} column density distribution
functions based on the 34 THINGS galaxy images with physical
resolution that varies between about 100 and 500~pc and no ability to
include opacity corrections. They compare these to the \citet{zwaa05}
curve by arbitrarily normalizing the distribution from each THINGS
galaxy to the Zwaan et al. curve at log(N$_{HI}$)~=~20, rather than
utilizing the space density of each target galaxy. Based on this
visual comparison they suggest that the \ion{H}{1} distribution
function is insensitive to physical resolution as coarse as several
kpc. Our own comparison, discussed above and shown in
Figs.~\ref{fig:fnhc} and \ref{fig:fnhcz}, utilizes an absolute
normalization and clearly demonstrates that both high physical
resolution and opacity corrections are vital in recovering the column
density distribution function accurately from images of \ion{H}{1}
21~cm emission.

The filled circles with error bars in Figure~\ref{fig:fnhc} represent the
$<z>~=~0.95$ QSO absorption line data of \citet{rao06}. The open
circles with error bars represent the high redshift QSO absorption
line data of \citet{proc09} and \citet{omea07}. Those results are
based on a sample of 738 damped Ly$\alpha$ absorption
systems (DLA, log(N$_{HI}$)~$>$~20.3) with $<z>~=~3.06$ and 78
``Super'' Lyman limit systems (SLLS, 19 $\le$ log(N$_{HI}$) $\le$ 20.3)
with $<z>~=~3.1$. The $<z>~=~3.6$ Lyman limit system data (LLS,
17.5 $\le$ log(N$_{HI}$) $\le$ 19) from \citet{proc10} is plotted as the
set of dot-long-dash lines. These parameterised fits, taken from
Table~5 of \citet{proc10}, represent the range of allowed values. (We
note in passing a minor typographical error in one of the tabulated
parameters, the constant log $k_{LLS}$ for one of these curves should
likely read $-3.1$ and not $-4.1$.) While the QSO absorption line data
is in good agreement with our z~=~0 distribution function for
log(N$_{HI}$)~=~21 to 21.5, there is an interesting trend apparent
for 18 $<$ log(N$_{HI}$) $<$ 21, in which the areal coverage of such
intermediate column density neutral gas declines continuously from
z~=~3 to z~=~0.

Another way of viewing all of the data presented above is to consider
the corresponding mass density as function of \ion{H}{1} column
density. We plot this distribution in the right hand panel of
Figure~\ref{fig:fnhc}. The mass density was calculated from,
\begin{equation}
\rho_{\rm HI}(N_{\rm HI}) = m_H {H_0 \over c} f(N_{\rm HI},X) N_{\rm HI}^2 {\rm ln}(10) d {\rm log}(N_{\rm HI})
\label{eqn:rnh}
\end{equation}
where $m_H$ is the mass of an \ion{H}{1} atom. This plot illustrates
the range of column density that dominates the contribution to mass
density. The trend noted previously for a declining contribution of
modest column density gas (18 $<$ log(N$_{HI}$) $<$ 21) with decreasing
redshift is clear. That decline corresponds to about a factor of 5
since z~=~3 at the DLA limit of log(N$_{HI}$)~$=$~20.3, and about a
factor of 2.5 relative to z~=~1. The z~=~3.6 LLS data demonstrate that
this excess neutral gas disappears by log(N$_{HI}$)~=~17.5, where the
areal coverage converges again with our z~=~0 distribution function.

The plots of Figure~\ref{fig:fnhc} are presented again over a
reduced range of column densities in Figure~\ref{fig:fnhcz} to permit a
more detailed comparison of the data and distributions, while the
numerical values are listed below in Appendix A for convenience. 

Since the primary observed difference between the z~=~0 and higher
redshift \ion{H}{1} distribution functions is limited to an increase
in the surface area of intermediate column density gas, we have
considered a toy model for the z~=~1 and z~=~3 images that simply
retains the smallest radii in the M31 image shown in
Figure~\ref{fig:m31nhc}, and replaces all structures at large
projected radii (greater than 25~kpc or 1.8 degree), with the apparent
column density version of the same image that is first made fainter by
a factor of three (z~=~1) and five (z~=~3) and scaled up in linear
size by a factor of 1.55 (z~=~1) and 2.3 (z~=~3). This could represent
an extended intermediate column density halo around the high column
density disk, that is larger in area than the actual z~=~0 galaxy by a
factor of 2.4 for z~=~1 and 5.1 for z~=~3. The resulting distribution
function and mass density function are overlaid on
Figs.~\ref{fig:fnhc} and \ref{fig:fnhcz} as the long-dashed and
short-dashed curves. These curves provide quite a good representation
of the QSO absorption line data at z~=~1 and z~=~3 of \citet{rao06},
\citet{proc09} and \citet{omea07}. 

While this toy model is not physically motivated, it
serves to demonstrate the types of structures needed to
account for the enhanced contributions of intermediate column density
gas in the high redshift data, and achieves these with only minimal
changes to the z~=~0 images. Such a distribution might arise if
gas accretion rates from the intergalactic medium were substantially
enhanced in the past. The recent simulations of \citet{kere09} suggest
that the total gas accretion rate was enhanced by factors of about 3
and 10 at redshifts of 1 and 3 over that at z~=~0, with the
enhancement of ``cold mode'' accretion of mildly ionized, 10$^{4-5}$~K
gas, being even more dramatic at these redshifts.

An alternative model, that is perhaps more plausible, might consist of
a population of dwarf satellites surrounding each massive high
redshift galaxy, the majority of which will merge with the parent by
z~=~0. Since the spatial concentration of such a system will be
similar to the halo model and the areal coverage of both high and
intermediate column density gas will need to be identical (to match
the QSO absorption line data) there will likely be only minor
practical differences between these two scenarios for the purpose of a
statistical description. We have not attempted to pursue modeling of a
merging satellite population at this time because of the many
additional free parameters that would be introduced.

\subsection{\ion{H}{1} pathlength functions}

\begin{figure*}
     \centering
     \includegraphics[trim=25 0 75 0,width=5.cm]{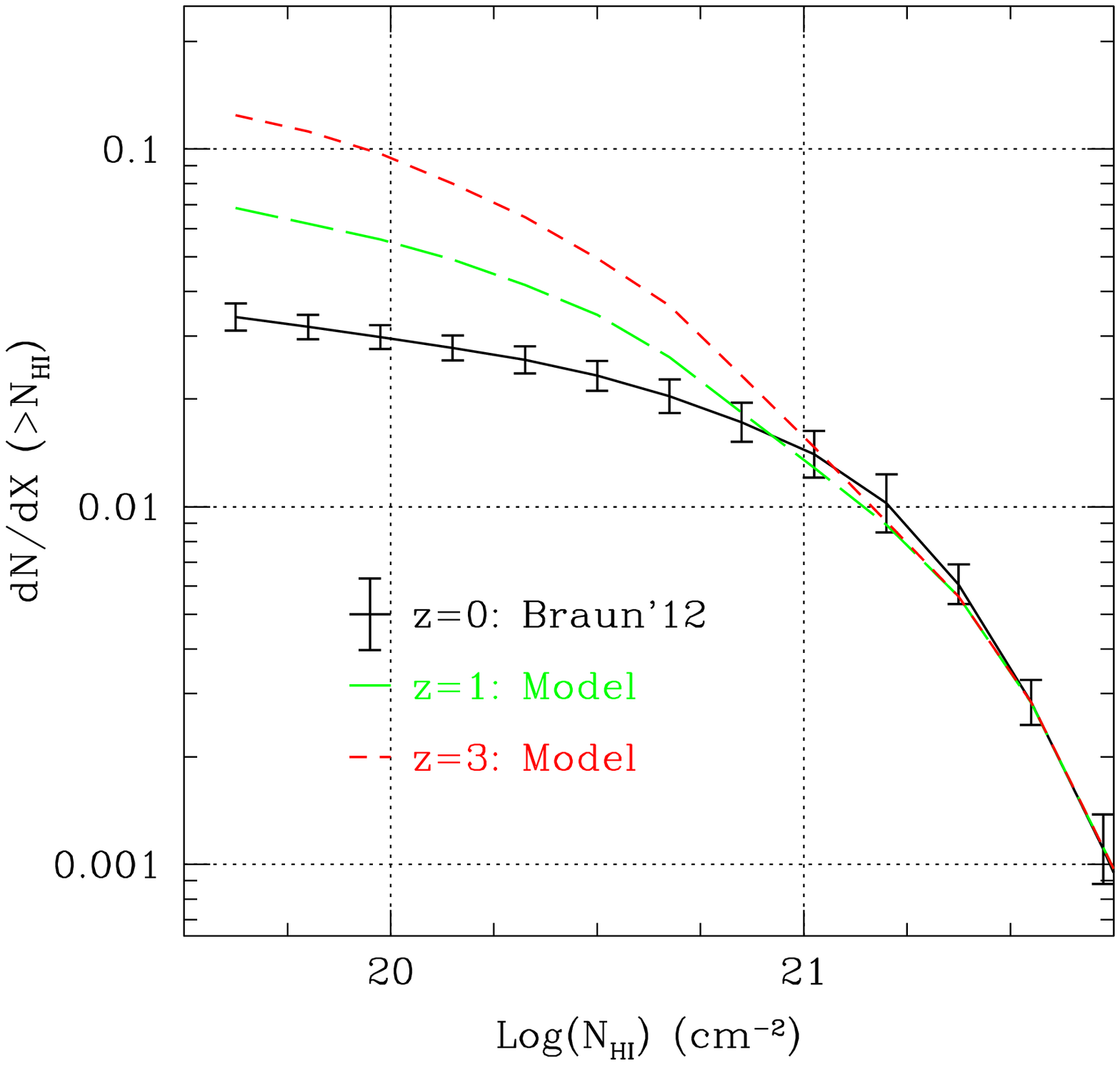}  
     \hfill
     \includegraphics[trim=50 0 50 0,width=5.cm]{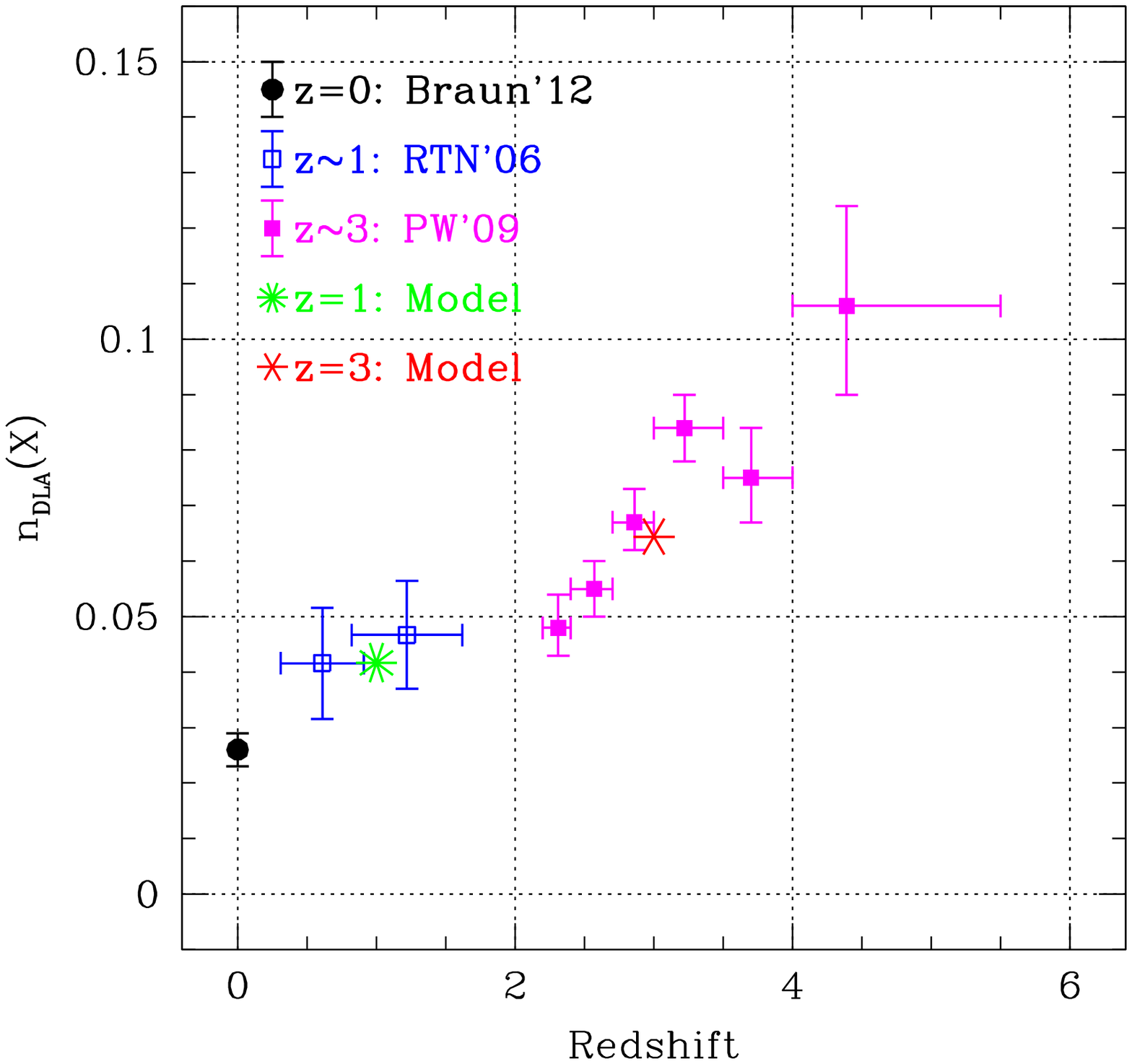}  
     \hfill
     \includegraphics[trim=75 0 25 0,width=5.cm]{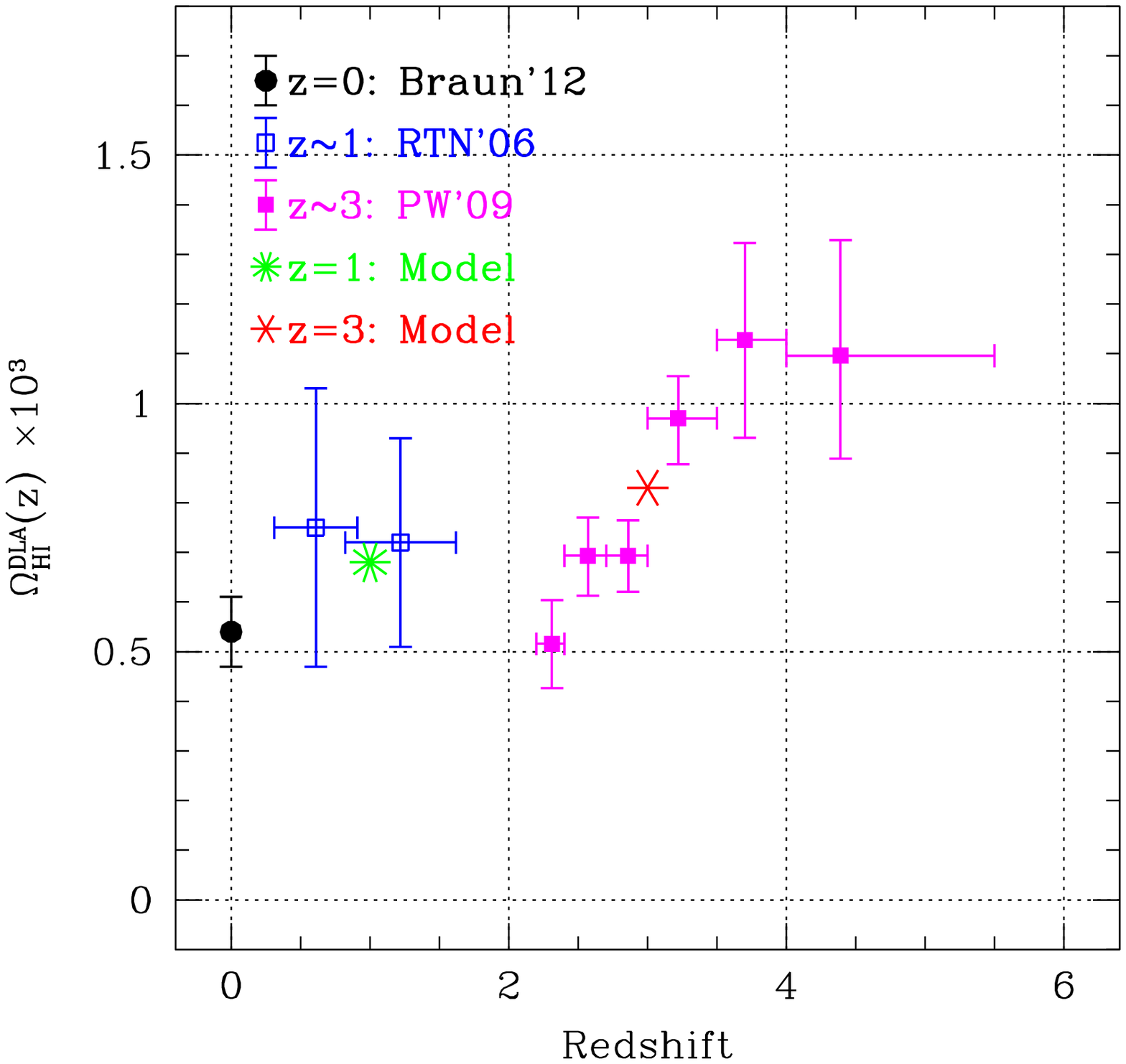}  
     \caption{Pathlength function of \ion{H}{1} column density (left),
       as well as redshift evolution of the total number (center) and
       \ion{H}{1} mass density (right) of DLA systems. }
     \label{fig:Nnh}
   \end{figure*}

A quantity that is often more useful in planning and interpreting
surveys is the pathlength function, $dN/dX$, the number of intervening
systems per unit co-moving distance along any line-of-sight that
exceed a particular column density. This is formed by integrating the
positive tail of the \ion{H}{1} distribution function and is shown in
Figure~\ref{fig:Nnh}. The two model curves discussed above to represent
the z~=~1 and 3 QSO absorption line data are also plotted in the
figure. The total number of equivalent Damped Lyman Alpha (DLA)
absorption systems, defined to be those with \ion{H}{1} column
density, $log(N_{HI}) > 20.3$, is $n_{DLA}(X,z=0)$~=~0.026$\pm$0.003
for our redshift zero determination. This is almost a factor of two
lower than the value estimated by \citet{zwaa05},
$n_{DLA}(X,z=0)$~=~0.045$\pm$0.006, and seems a consequence of the
insufficient physical resolution (a median of 1.4~kpc relative to a
structure scale of 100~pc) of that study. Our measurement is compared
with the values of $n_{DLA}(X,z)$ published in the studies of
\citet{rao06} and \citet{proc09} in the center panel of
Figure~\ref{fig:Nnh}. There has been a continuous decline in
$n_{DLA}(X)$ with redshift as noted by many previous authors. Our
redshift zero measurement demonstrates an ongoing decline since z~=~1,
rather than flattening to a constant value as suggested by the
\citet{zwaa05} point. We include the $n_{DLA}(X,z)$ values of our
model for the z~=~1 ($n_{DLA}(X,z=1)$~=~0.042) and z~=~3
($n_{DLA}(X,z=3)$~=~0.064) distributions on the plot to demonstrate
the degree to which they are in agreement with the published data. The
right-hand panel of Figure~\ref{fig:Nnh} presents the redshift evolution
of $\Omega_{HI}^{DLA}$, the mass density in \ion{H}{1} due to systems
with column densities $log(N_{HI}) > 20.3$. This was calculated for
redshift zero by considering the integral mass above this threshold,
which amounts to 96\% of the total \ion{H}{1} mass in galaxies or
$\Omega_{HI}^{DLA}(z=0)$~=~5.4$\pm0.9 \times 10^{-4}$. The data points
in the figure are those of \citet{rao06} (after rescaling to an
\ion{H}{1} only value) and of \citet{proc09}. For comparison we also plot
the mass densities in our redshifted model distributions to
demonstrate their consistency.

\subsection{Relating 21~cm absorption to \ion{H}{1} column density}

The 21~cm \ion{H}{1} transition offers great promise for documenting
the evolution of neutral gas content with cosmic time. However, there
are specific issues which need to be overcome before it can fulfill
that promise. In the case of 21~cm emission, the issue is
sensitivity. Even the deepest integrations with existing facilities
have only been able to push back this frontier to z~$\sim$~0.2 for
individual target detection \citep{verh10}. And while so-called
``stacking'' analysis \citep{lah07} can provide some statistical
information on gas content at larger distances, the interpretation of
such results can not be disentangled from the definition of the sample
of objects to be co-added. Reaching down to below M$_*$ on
individual targets at redshifts exceeding unity will require about 100
times the sensitivity that is currently available, which can only be
achieved with a hundred-fold increase in collecting area such as
planned for Phase 2 of the international Square Kilometre Array
\citep{cari04}. For 21~cm line absorption, the major limitations are
firstly, the need for a large instantaneous field-of-view which would
enable effective blind surveys toward background continuum sources and
secondly, an effective calibration of integrated absorber strength
with the associated \ion{H}{1} column density. The limitation of
field-of-view is currently being addressed with Phased Array Feed
(PAF) technology deployed on parabolic reflectors \citep{debo09} that
will provide as much as 30~deg$^2$ of instantaneous sky coverage. The
limitation of 21~cm absorber calibration is one that we will attempt
to address here.

\begin{figure*}
     \centering
     \includegraphics[trim=0 0 50 0,width=8.cm]{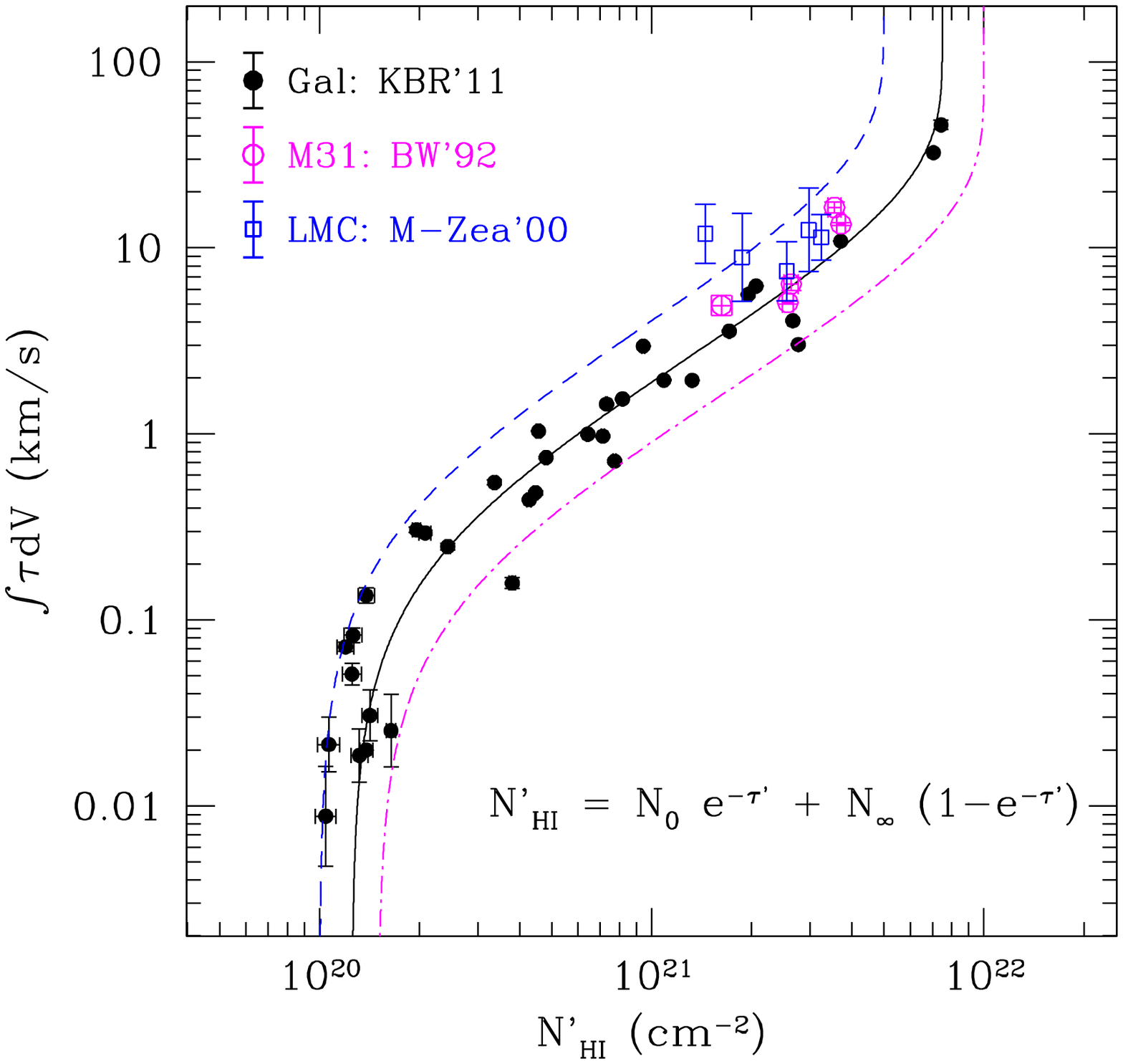}  
     \hfill
     \includegraphics[trim=50 0 0 0,width=8.cm]{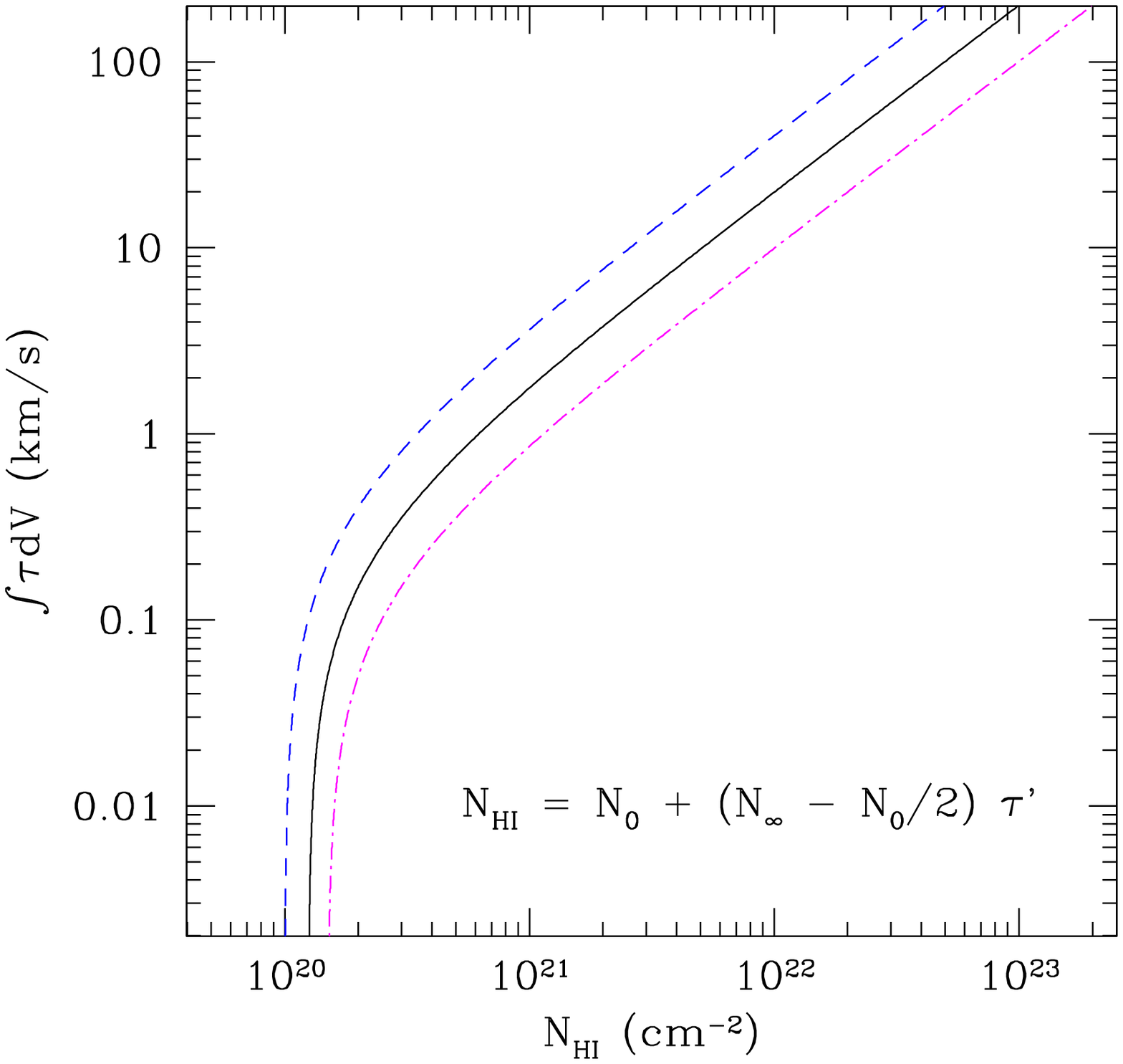}  
     \caption{The relation of integrated 21~cm absorption with apparent
       neutral hydrogen column density (left) and total neutral
       hydrogen column density (right).}
     \label{fig:nhatau}
   \end{figure*}

The necessity for calibration of the 21~cm absorption line strength is
determined by the intrinsic temperature sensitivity of the 21~cm
\ion{H}{1} line opacity, which varies inversely with temperature, as
already noted in eqn.~\ref{eqn:tau} as well as on possible nonthermal
contributions to the linewidth. Since any particular line-of-sight is
likely to traverse more than one temperature of medium, there might be
no simple correspondence of the 21~cm line opacity with the underlying
\ion{H}{1} column density. Fortunately, there appears to be a
particularly good one-to-one correspondence of these quantities
observationally. \citet{kane11} have recently demonstrated the
remarkably tight relationship between integrated 21~cm absorption
opacity and the apparent (ie. uncorrected for opacity) column density
determined from complimentary 21~cm emission observations of
neighbouring lines-of-sight. We present a similar plot in
Figure~\ref{fig:nhatau} where the Galactic 21~cm absorption and emission
data of \citet{kane11} are plotted as filled circles, together with
similar data obtained toward M31 by \citet{brau92} as open circles and
the LMC by \citet{marx00} as open squares. Only the high
signal-to-noise absorption detections (exceeding 4$\sigma$) are
plotted. The curves in the figure represents the expectation for a
``sandwich'' geometry of cool gas with properties ($T_c, \tau_c$)
surrounded by layers of warm gas with properties ($T_w, \tau_w$) as
discussed in \citet{kane11} from,
\begin{equation}
N'_{HI} = N_{0} e^{-\tau_c'} + N_{\infty} (1-e^{-\tau_c'}) 
\label{eqn:nha}
\end{equation}
for a measured ``apparent'' column density, $N'_{\rm HI}$ (based on
the integral of observed brightness temperature), a threshold column
density (where $\tau_c \rightarrow 0$), $N_0 \sim T_w \tau_w \Delta
V$, a saturation column density (where $\tau_c \rightarrow \infty$),
$N_\infty \sim (T_c+T_w \tau_w / 2)\Delta V$ and an effective opacity,
$\tau_c'$. The effective opacity is related to the measured integrated
opacity by the effective linewidth, $\Delta V$, as $\int \tau {\rm dV}
= \tau_c' \Delta V$. The solid curve is for (${\rm N_{0}} = 1.25
\times 10^{20}$~\cm, ${\rm N_{\infty}} = 7.5 \times 10^{21}$~\cm,
$\Delta V = 15$~km/s) while the dashed and dot-dash curves are for
(${\rm N_{0}} = 10^{20}$~\cm, ${\rm N_{\infty}} = 5.0 \times
10^{21}$~\cm, $\Delta V = 20$~km/s) and (${\rm N_{0}} = 2 \times
10^{20}$~\cm, ${\rm N_{\infty}} = 10^{22}$~\cm, $\Delta V = 10$~km/s),
respectively.

The ``sandwich'' geometry is meant to represent the role of a
``temperature-shielding'' column density of the Warm Neutral Medium
(WNM) that absorbs local EUV and X-ray radiation and is a requirement
for condensation of the Cool Neutral Medium (CNM) within
\citep[e.g.][]{wolf03}. It should be noted that this
``temperature-shielding'' column is quite distinct from the
``self-shielding'' column density of only a few times $10^{17}$~\cm~
that marks the onset of recombination of hydrogen within an ionizing
radiation field \citep[e.g.][]{dove94}. Once neutral columns of about
$10^{19}$~\cm~are achieved, the expectation is that the ionization
fraction will be minimal, even in extreme radiation fields. Such a
model provides a very good characterisation of the observable
properties of the absorbing medium in the Galaxy, where it is best
constrained by data, but also of the available extragalactic data from
M31 and the LMC, in which both higher and lower metallicity and
radiation field environments are probed. The dashed and dot-dash
curves provide a reasonable estimate of the scatter observed in this
relationship. Since the curves drawn in Figure~\ref{fig:nhatau} are
based on a physical model, we can also write the corresponding
equation for the {\it actual rather than the apparent} \ion{H}{1}
column density,
\begin{equation}
N_{HI} = N_{0} + (N_{\infty} - {N_{0} \over 2}) \tau_c'
\label{eqn:nh}
\end{equation}
The curves based on the same parameters as previously are drawn in the
right-hand panel of Figure~\ref{fig:nhatau}. While these are unchanged
at low column density, they do not exhibit the saturation effect at
high columns that afflict observations of 21~cm \ion{H}{1} emission.

\subsection{21~cm opacity distribution function}

\begin{figure*}
   \centering
   \includegraphics[trim=0 0 50 0,width=8.cm]{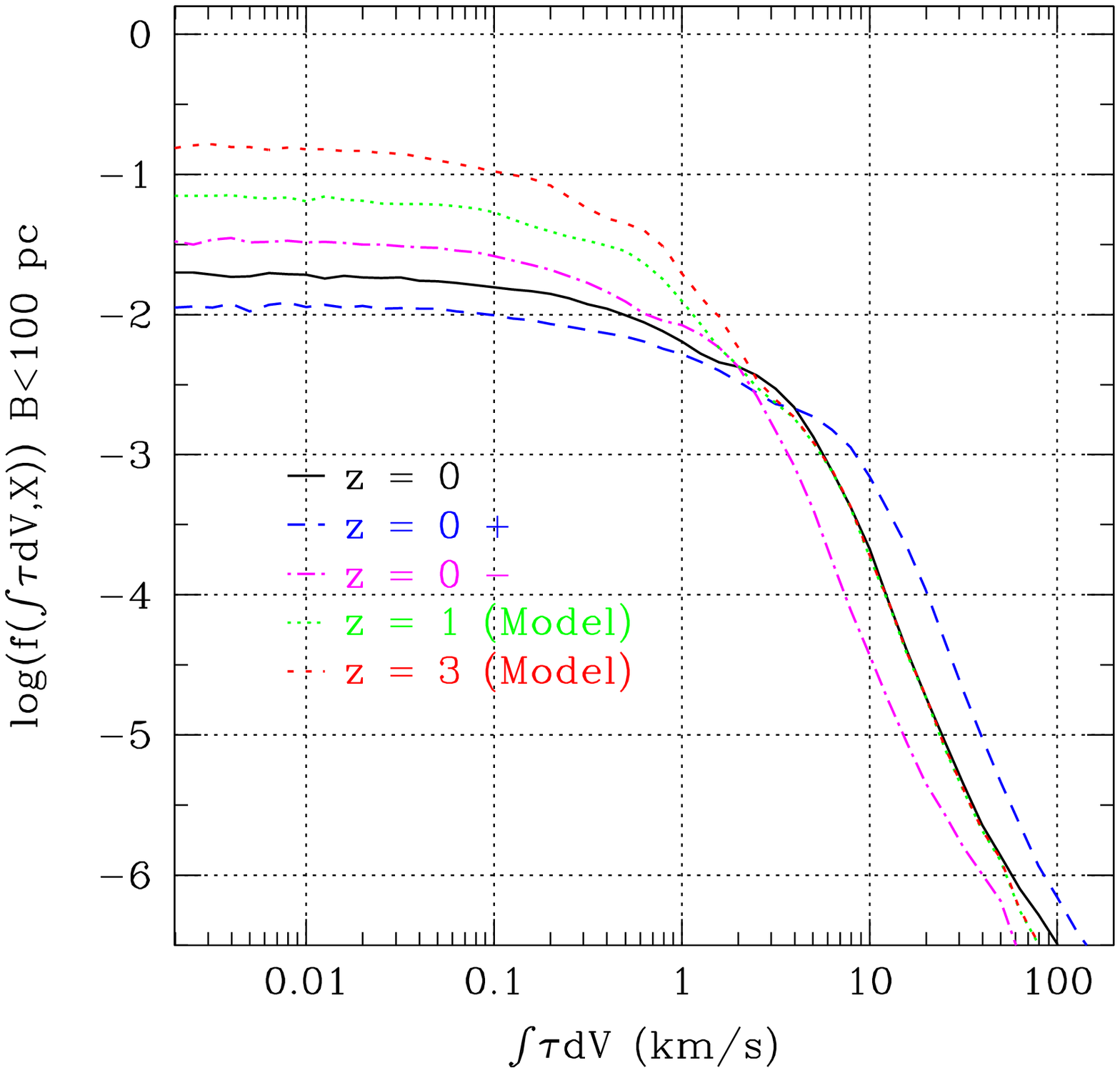}
   \hfill
   \includegraphics[trim=50 0 0 0,width=8.cm]{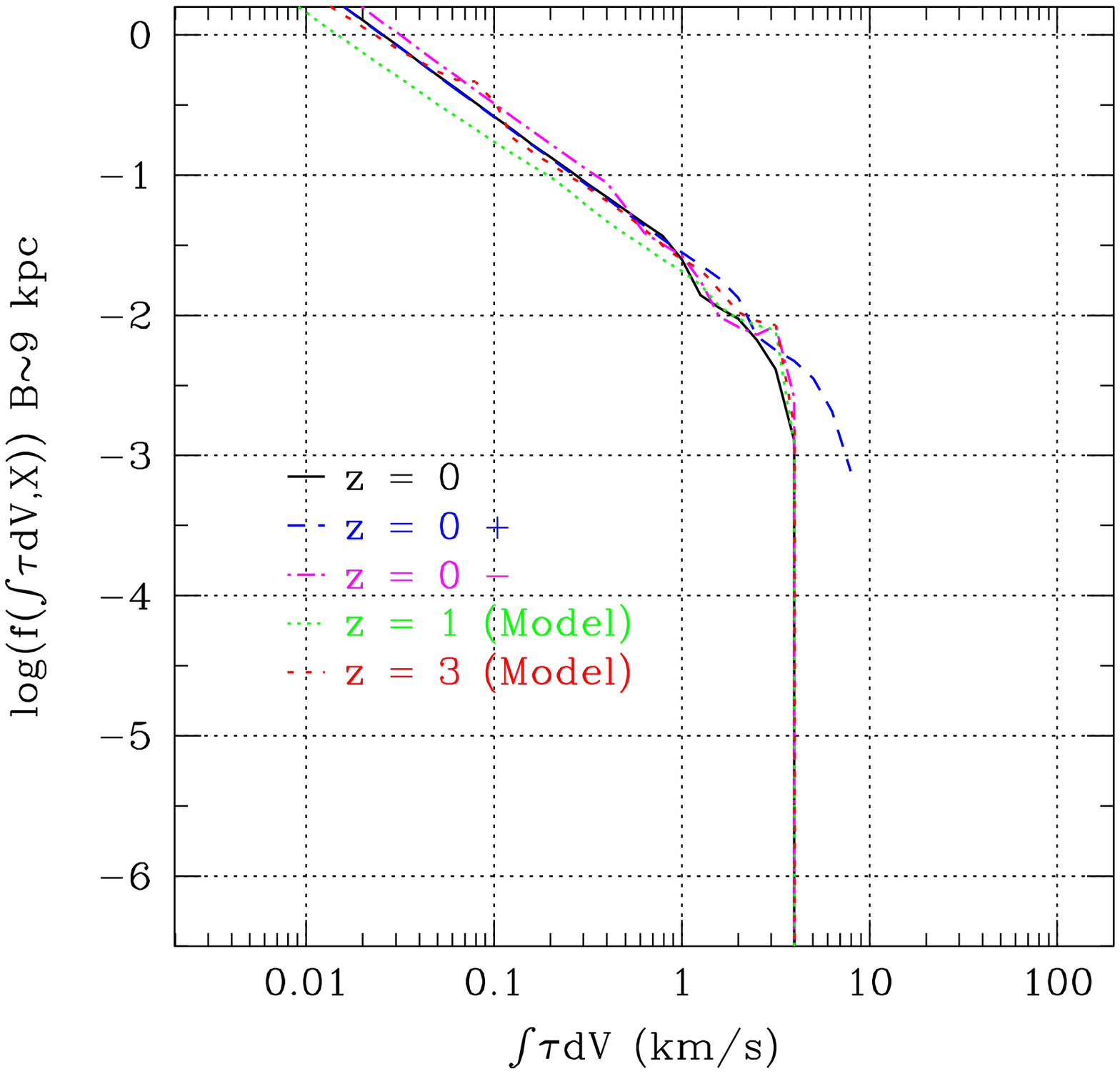}  
   \caption{Distribution function of integrated 21~cm absorption at
     full resolution (left) and observed with 9 kpc effective
     resolution (right).}
   \label{fig:ftau}
   \end{figure*}

With the column density calibration of 21~cm absorption line strength
of the previous sub-section in hand it becomes possible to construct
the distribution function of 21~cm opacity. Eqn.~\ref{eqn:nh} was used
to transform the three images of opacity-corrected column density
(Figs.~\ref{fig:m31nhc}--\ref{fig:lmcnhc}) to corresponding images of
21~cm absorption line strength. The transformation was done using the
nominal transformation depicted in Figure~\ref{fig:nhatau} with
parameters (${\rm N_{0}} = 1.25 \times 10^{20}$~\cm, ${\rm N_{\infty}}
= 7.5 \times 10^{21}$~\cm, $\Delta V = 15$~km/s) as well as the two
envelope curves with (${\rm N_{0}} = 10^{20}$~\cm, ${\rm N_{\infty}} =
5.0 \times 10^{21}$~\cm, $\Delta V = 20$~km/s) and (${\rm N_{0}} = 2
\times 10^{20}$~\cm, ${\rm N_{\infty}} = 10^{22}$~\cm, $\Delta V =
10$~km/s) in order to track the intrinsic scatter and uncertainty. 
The distribution function was then calculated from eqn.~\ref{eqn:fnhc}
but with $\int \tau dV$ in the place of ${\rm N_{HI}}$.
This is shown in Figure~\ref{fig:ftau} with
the solid curve illustrating the nominal z~=~0 relation, while the
dashed and dot-dashed curves span the scatter and
uncertainty. What is notable is the flatness of the distribution below
about $\int \tau dV = 0.1$~\kms. The ``temperature-shielding''
threshold implies that there will be essentially no additional systems
seen with an intrinsic absorption depth below this value.

Distribution functions of the 21~cm absorption for the toy models
developed above to describe the z~=~1 and 3 QSO data are also plotted
in Figure~\ref{fig:ftau} as the dotted and short-dashed curves utilizing the
nominal z~=~0 transformation. There is considerable uncertainty in the
applicability of the z~=~0 relation at these higher redshifts. The
intergalactic ionizing radiation field is expected to increase
dramatically with redshift in concert with the evolution of the
massive star formation rate and active galaxy incidence. While this
will have a major influence on the ``self-shielding'' column density
(of a few times $10^{17}$~\cm) for the onset of the ionized to neutral
transition at the galaxy-IGM interface, it is less clear what the
impact might be on the ``temperature-shielding'' column (of $\sim
10^{20}$~\cm) for the warm to cool transition that occurs deep within
individual galaxy disks. This is because the radiation field in this
latter case is strongly dominated by local contributions
\citep[e.g.][]{wolf03}. It is likely that the intrinsic scatter of the
z~=~0 relation in Figure~\ref{fig:nhatau} is already a consequence of
the large variations in local radiation field intensity that naturally
occur within individual galaxies. Future work will be vital to test
this hypothesis in more extreme environments, such as the disks of
nearby star-bursting galaxies. We will continue with our discussion of
possible evolutionary trends of absorption properties with this caveat
in mind.

Since deep 21~cm absorption features, $\int \tau dV > 5$~\kms, require
high column density, ${\rm N_{HI}} > 10^{21.3}$~\cm, lines-of-sight
and the incidence of such column densities does not appear to evolve
significantly with redshift (Figure~\ref{fig:fnhcz}), there is also no
evolution in the distribution function of 21~cm absorption in
Figure~\ref{fig:ftau} for such deep features. Redshift evolution is
confined to the relatively weak 21~cm absorbers with strength $\int
\tau dV < 1$~\kms~ that correspond to ${\rm N_{HI}} < 10^{21}$~\cm.

   \begin{figure*}
   \centering
   \includegraphics[trim=0 0 50 0,width=8.cm]{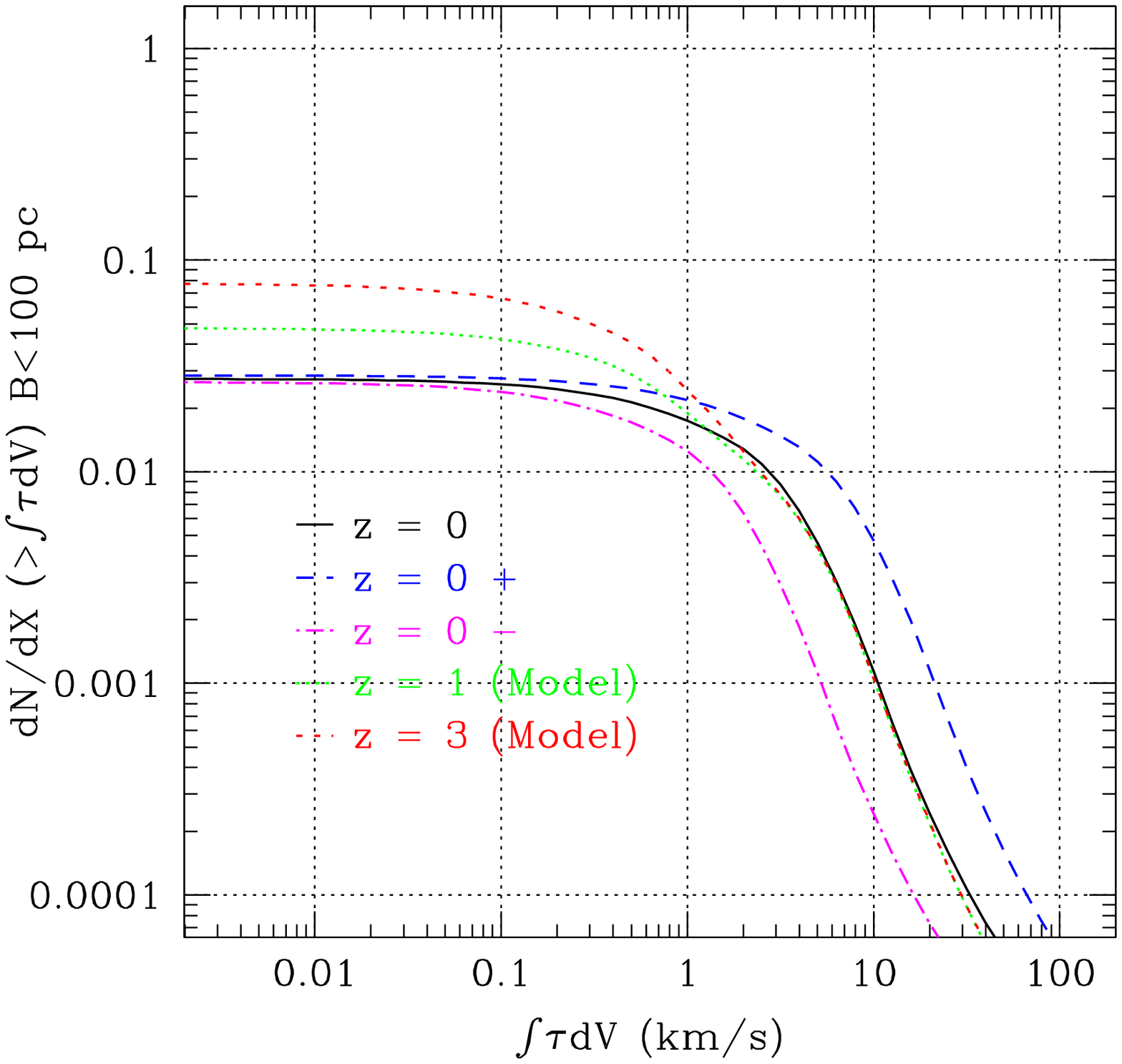}
   \hfill
   \includegraphics[trim=50 0 0 0,width=8.cm]{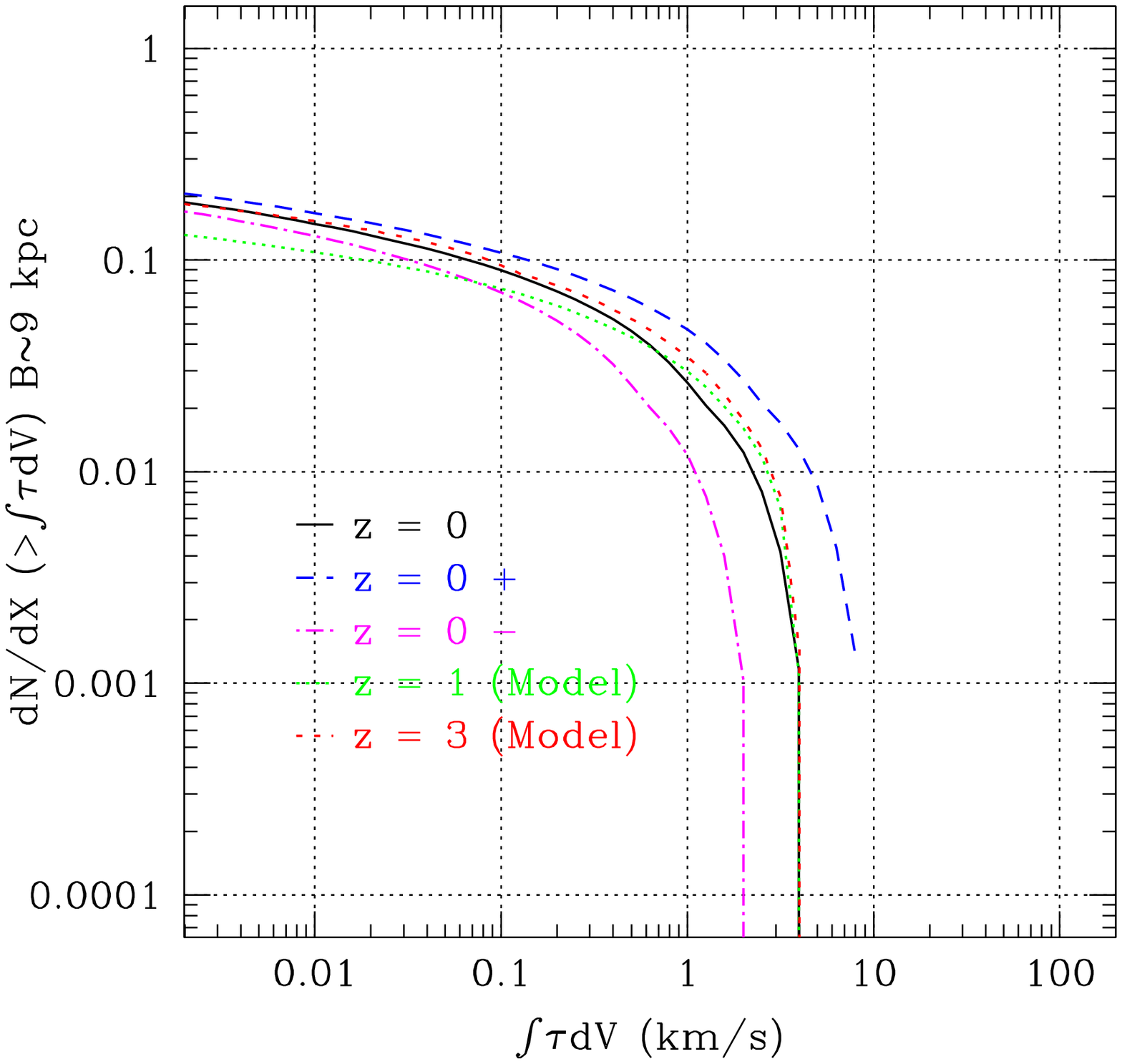}  
   \caption{Pathlength function of integrated 21~cm absorption observed
     at full resolution (left) and observed with 9 kpc effective
     resolution (right).}
   \label{fig:Ntau}
   \end{figure*}

\subsection{The pathlength function of 21~cm absorption}

Just as was done above for the \ion{H}{1} column density, it is
possible to calculate the corresponding pathlength function, $dN/dX$,
the number of 21~cm absorption systems per unit distance that exceed a
particular absorption depth, $\int \tau dV$. This is formed by
integrating the positive tail of the 21~cm absorption distribution
function and is shown in Figure~\ref{fig:Ntau} with the same curves used
previously to depict the z~=~0, 1 and 3 predictions. The same trends
noted previously for the distribution function apply here as well,
namely an absence of redshift evolution for deep absorbers and number
evolution by a factor of several for the faint absorbers.

\subsection{The impact of finite background source size}

The preceeding discussion has assumed that the background source used
to probe intervening absorbers has a small angular size relative to
the most compact structures within the absorbing system. What if this
is not the case? Since we already have images that represent
absorption depth, it is straightforward to simulate the observable
effect of an arbitrary background source size and structure. This can
be done by evaluating the convolution of the absorption depth image
with an image of the background source, normalized to preserve the
integrated absorption depth. The distribution function of such
convolved images can then be calculated in the same way. This has
been done in the right hand panels of Figs.~\ref{fig:ftau} and
\ref{fig:Ntau} for an assumed Gaussian background source of 9~kpc
FWHM. This choice of background source model is motivated by the 1.3
arcsec median source size observed for mJy brightness radio sources at
GHz frequencies \citep{wind03} which subtend more than 9~kpc for all
redshifts greater than about 0.7. This should be regarded as a ``worst
case scenario'' from the standpoint of absorption signal dilution,
since there is some probability that a compact component may be
present, even in an intrinsically extended background source.

The impact of a significantly extended background source on the 21~cm
absorber distribution and pathlength functions is dramatic. As can be
seen in the right hand panels of Figs.~\ref{fig:ftau} and \ref{fig:Ntau},
deep absorption systems will not be observed at all in this instance,
while an enhanced number of apparently faint absorbers are seen
instead. This redistribution of intrinsically deep into apparently
faint absorbers also has as a consequence that there is no predicted
observable evolution at all in the 21~cm absorption properties with
redshift.

As noted above, this assessment must be regarded as a ``worst case
scenario'' since the median source size has been distributed smoothly
in our simulation. Better insights into the typical structures seen in
radio sources at GHz frequencies are provided by the recent study of
\citet{midd11} who studied a sample of nearly 100 moderate brightness
radio sources with angular resolution of about 20~mas. Overall, some
20\% of these sources were found to have a substantial (50 -- 100\%)
flux contribution from unresolved structures. However, when
considering sources fainter than 1~mJy at 1.4~GHz, this fraction
declined to only 7\%. 

What the dramatic contrast between the left- and right-hand panels of
Figs.~\ref{fig:ftau} and \ref{fig:Ntau} underlines is the importance
of isolating the flux contribution of suitably compact structures,
subtending less than about 100~pc at the distance of the absorption
system, in interpreting any measurement of 21~cm
absorption. Interpretation is essentially impossible without this
information. The implication for large surveys of \ion{H}{1} 21~cm
absorption at redshifts z~$>$~0.5 is that high resolution imaging,
with about 15~mas beam size, will be needed for the useful
interpretation of any detections, either during the survey itself or
as follow-up. At the relevant observing frequencies,
$\nu$~$<$~700~MHz, this implies a need for baselines greater than
about 5500~km.

\section{Conclusions}
\label{sec:conc}

The analysis of high resolution and sensitivity \ion{H}{1} datacubes
of Local Group galaxies has permitted some fundamental conclusions to
be drawn:

\begin{enumerate}

\item Galaxy disks contain a significant population of 
  \ion{H}{1} features which are self-opaque in the 21~cm
  transition. Such features are intrinsically compact, with narrow
  dimensions that are typically only about 100~pc, which determines
  the resolution required for their study. Such features can locally
  disguise the presence of high atomic column densities in the range
  $22 < log(N_{HI}) < 23$, that are not apparent in images of the
  integrated 21~cm linestrength. Globally, such features account for
  about 34$\pm$5\% more atomic gas than is derived under the assumption of
  negligible self-opacity.

\item Application of this opacity correction to recent estimates of
  the cosmological mass density of \ion{H}{1} within local galaxies yields,
  $\Omega_{HI}^{gal}(z=0)$~=~5.6$\pm0.9 \times 10^{-4}$ assuming that
  our limited galaxy sample provides a representative determination. 

\item The resolved, opacity corrected images of Local Group galaxies
  permit calculation of a robust redshift zero \ion{H}{1} distribution
  function, $f(N_{HI},X)$, with excellent statistical properties
  (based on 1000's of independent sight-lines) and a high degree of
  internal consistency amongst galaxy types between Sb and Sm. While similar
  to high redshift determinations, there is evidence for a systematic
  decline in intermediate column density gas, $18 < log(N_{HI}) <
  21$, that can be well modelled by a decrease in the surface area of
  galaxy halos (or associated satellites) by a factor of about 2.5 since z~=~1
  and a factor of about 5 since z~=~3.

\item The number of equivalent Damped Lyman Alpha (DLA) \ion{H}{1}
  absorbers per unit comoving length,
  $n_{DLA}(X,z=0)$~=~0.026$\pm$0.003, is significantly lower than seen
  at higher redshifts by factors as large as about four and follows a
  smoothly declining trend. Similarly, the cosmological mass density
  in DLAs, $\Omega_{HI}^{DLA}(z=0)$~=~5.4$\pm0.9 \times 10^{-4}$,
  follows a smoothly declining trend, albeit one that sees only a
  factor of two decline between redshift 4 and the present.

\item There is a tight, non-linear correlation between the observed 21~cm
  absorption opacity and the integrated 21~cm emission brightness from
  adjacent sight-lines through nearby galaxy disks which permits
  ``calibration'' of our \ion{H}{1} column density images into
  corresponding ones of integrated 21~cm opacity.

\item We present predictions for the distribution and pathlength
  functions of 21~cm integrated opacity at both redshift zero and at
  high redshift for the case of no evolution in the local opacity
  calibration. The prediction is that faint 21~cm absorbers, $\int\tau
  dV < 1$~\kms, have declined by a factor of about 2.5
  since z~=~1 and a factor of about 5 since z~=~3, while the number of deep
  absorbers, $\int\tau dV > 5$~\kms, has not evolved.

\item We explore the impact of a finite background source size on the
  distribution and pathlength functions of 21~cm integrated
  opacity. Adoption of the observed 1.3 arcsec median source size of
  mJy radio sources dramatically changes the apparent opacity
  distributions and effectively precludes any physical interpretation
  of the measurements. Future surveys of 21~cm absorption will require
  a 15 mas beamsize for their unambiguous interpretation.

\end{enumerate}


%% Included in this acknowledgments section are examples of the
%% AASTeX hypertext markup commands. Use \url without the optional [HREF]
%% argument when you want to print the url directly in the text. Otherwise,
%% use either \url or \anchor, with the HREF as the first argument and the
%% text to be printed in the second.

\acknowledgments

We acknowledge the useful comments of Jason Xavier Prochaska, Elaine
Sadler and Dave Thilker on an earlier version of this manuscript. The Westerbork
Synthesis Radio Telescope is operated by ASTRON (Netherlands
Foundation for Research in Astronomy) with support from the
Netherlands Foundation for Scientific Research (NWO). The National
Radio Astronomy Observatory, which operates the Green Bank Telescope
and Very Large Array, is a facility of the National Science Foundation
operated under cooperative agreement by Associated Universities,
Inc. The Australia Telescope and the Parkes are funded by the
Commonwealth of Australia for operation as a National Facility managed
by CSIRO.


%% in parentheses, after the keyword, but they will not be verified.

{\it Facilities:} \facility{WSRT}, \facility{VLA}, \facility{ATCA}, \facility{Parkes}, \facility{GBT}.



\appendix
\section{Numerical values of the redshift zero \ion{H}{1} distribution function}

The numerical values of the \ion{H}{1} distribution function presented
in this paper are tabulated below. The data source for the different
column density ranges are indicated. The error estimates in the third
column are determined from the number of independant image pixels used
in the evaluation for the data of references 1 and 2, while for
reference 3 this is determined from the RMS deviation amongst the
three contributing galaxies, which exceeds the statistical error
estimate by factors of between four and 20.

\begin{deluxetable}{cccc}
\tablewidth{0pt}
\tablecaption{\ion{H}{1} distribution function at z~=~0}
\tablehead{
\colhead{$log(N_{HI})$[\cm]}           & \colhead{$log(f(N_{HI},X)$}      &
\colhead{$\Delta log(f(N_{HI},X)$}   & \colhead{Ref} }      
\startdata
       16.75  &    -17.28  &    0.093  & 1 \\
       17.00  &    -17.43  &    0.083  & 1 \\
       17.25  &    -17.74  &    0.089  & 1 \\
       17.50  &    -18.38  &    0.140  & 1 \\
       17.75  &    -18.82  &    0.174  & 1 \\
       18.00  &    -19.25  &    0.213  & 1 \\
       18.25  &    -19.69  &    0.265  & 1 \\
       18.50  &    -20.14  &    0.334  & 1 \\
       18.75  &    -20.44  &    0.353  & 1 \\
\hline
       18.58  &    -20.30  &    0.188  & 2 \\
       18.84  &    -20.60  &    0.195  & 2 \\
       19.11  &    -21.06  &    0.246  & 2 \\
       19.37  &    -21.40  &    0.269  & 2 \\
       19.63  &    -21.82  &    0.322  & 2 \\
       19.89  &    -22.15  &    0.348  & 2 \\
\hline
       19.62  &    -21.91  &    0.321  & 3 \\ 
       19.80  &    -22.10  &    0.227  & 3 \\ 
       19.97  &    -22.29  &    0.121  & 3 \\
       20.15  &    -22.43  &    0.027  & 3 \\
       20.32  &    -22.53  &    0.032  & 3 \\
       20.50  &    -22.64  &    0.058  & 3 \\
       20.67  &    -22.79  &    0.067  & 3 \\
       20.85  &    -22.95  &    0.044  & 3 \\
       21.02  &    -23.05  &    0.109  & 3 \\
       21.20  &    -23.19  &    0.182  & 3 \\
       21.37  &    -23.47  &    0.086  & 3 \\
       21.55  &    -23.92  &    0.084  & 3 \\
       21.73  &    -24.47  &    0.139  & 3 \\
       21.90  &    -25.16  &    0.161  & 3 \\
       22.07  &    -25.76  &    0.195  & 3 \\
       22.25  &    -26.30  &    0.152  & 3 \\
       22.43  &    -26.73  &    0.161  & 3 \\
       22.60  &    -27.09  &    0.315  & 3 \\
       22.77  &    -27.52  &    0.332  & 3 \\
       22.95  &    -28.25  &    0.461  & 3 \\
\enddata
\tablerefs{
(1) Braun \& Thilker 2004; (2) Thilker et al. 2004; (3) Current study}
\end{deluxetable}




\clearpage

\end{document}